\documentclass[preprint,showpacs,floatfix,pre]{revtex4}
\usepackage[dvips]{graphicx}
\usepackage{dcolumn}
\usepackage{amsfonts}
\usepackage{amsmath}
\usepackage{amssymb}

\begin{document}

\title{The atomic approach of the Anderson model for the U finite case:
application to a quantum dot}
\author{T. Lobo}
\author{M. S. Figueira}
\email{figueira@if.uff.br}
\affiliation{Instituto de F\'{\i}sica - Universidade Federal Fluminense, \\
Av. Litor\^{a}nea s/n, 24210-340 Niter\'oi-RJ, Brazil }
\author{M. E. Foglio}
\affiliation{Instituto de F\'{\i}sica Gleb Wataghin - Universidade Estadual de Campinas, 
\\
Bar\~ao Geraldo 13083-970 Campinas-SP, Brazil}
\date{\today}

\begin{abstract}
In the present work we apply the atomic approach to the single impurity
Anderson model (SIAM). A general formulation of this approach, that can be
applied both to the impurity and to the lattice Anderson Hamiltonian, was
developed in a previous work (arXiv:0903.0139v1 [cond-mat.str-el]). The
method starts from the cumulant expansion of the periodic Anderson model
(PAM), employing the hybridization as perturbation. The atomic Anderson
limit is analytically solved and its sixteen eigenenergies and eigenstates
are obtained. This atomic Anderson solution, which we call the (AAS), has
all the fundamental excitations that generate the Kondo effect, and in the
atomic approach is employed as a \textquotedblleft seed \textquotedblright\
to generate the approximate solutions for finite $U$. The width of the
conduction band is reduced to zero in the AAS, and we choose its position so
that the Friedel sum rule (FSR) be satisfied, close to the chemical
potential $\mu$.

We perform a complete study of the density of states of the SIAM in all the
relevant range of parameters: the empty dot, the intermediate valence
(IV-regime),the Kondo and the magnetic regime. In the Kondo regime we obtain
a density of states that characterizes well the structure of the Kondo peak.
To shown the usefulness of the method we have calculated the conductance of
a quantum dot, side coupled to a conduction band.
\end{abstract}

\pacs{36.20.Ey,64.60.Cn,05.50.+q}
\maketitle

\section{Introduction}

\label{sec1}

Experimental results indicate \cite{Clogston} that localized magnetic
moments could appear when magnetic impurities are dissolved in a metal, and
a minimum in the electrical resistivity is always present in the material at
low temperatures $T$. Jun Kondo was the first who associated the formation
of magnetic moments in the material with the minimum of the resistivity. He
studied the phenomena employing perturbation theory in second order in the
Born approximation, \cite{Kondo64} and showed that the minimum was generated
by the spin-flip scattering of the conduction electrons with the magnetic
moment of the impurity; this phenomena become known as the Kondo effect. In
its calculation Kondo showed that the resistivity increases with the
decrease of temperature according to $ln(T)$, which become known as the
logarithmic divergence of the Kondo effect. In the alloys formed in such a
way it is not possible to control the Kondo parameters microscopically, but
with the advent of nanotechnology the Kondo effect was realized
experimentally in quantum dots (QD), with complete control over all the
relevant parameters \cite{Goldhaber98}.

Quantum dots are artificial atoms that offer a high degree of parameter
control by means of simple gate electrostatics, allowing the nanometric
confinement of the electrons. The coupling between the quantum dot and the
electronic reservoirs produces tunneling events that can modify the discrete
energy levels of the dot, changing them into complicated many-body wave
functions. When conducting electrons move in an out the nanostructure, a
progressive screening of the atomic spin occurs, in complete analogy with
the well-known Kondo effect in solids containing magnetic impurities. The
Kondo effect in quantum dots is then observed as a zero-bias conductance
resonance, associated with the entangled state of the electrons in the
electronic reservoirs and in the dot.

The experimental realization of the Kondo effect in QD's renewed the
interest in searching new approximate solutions for the single impurity
Anderson model (SIAM) and several calculations of the Green's functions (GF)
employing the method of motion equation (EOM) were reconsidered, \cite%
{Meir93,Luo99} but such methods suffered several drawbacks due to their
failure to satisfy the completeness and the Friedel sum rule. \cite%
{Completeness07} Another method that was extensively applied to describe the
SIAM is the slave boson mean field theory, \cite{Coleman1} but this method
suffered from a spurious second order phase transition. \cite{X-boson1}
Another recent approach that describes the SIAM is the local moment
approach, \cite{Logan98} that satisfies a number of correct limits:
weak-coupling Fermi liquid, atomic limit, and the Kondo regime, but the
strong coupling limit is plagued with a symmetry breaking and an unphysical
static order value of the local magnetic moment. Very recently appeared an
analytical calculation, \cite{Janis08} that employed diagrammatic expansions
with Feynman diagrams and obtained the Kondo resonance in the electron-hole
symmetric limit, but out of this limit the Kondo resonance was lost. Another
appealing approach is the quantum Monte Carlo technique, \cite{Monte_Carlo}
that is expensive from the computational point of view and is restricted
from intermediate to high temperatures. Other expensive computational
numerical methods are the dynamical mean field theory (DMFT) \cite{Kraut}
and the density-matrix renormalization group (DMRG). \cite{Schollwock2005}
The most effective approach which nowadays describes the dynamical
properties of the SIAM is the numerical renormalization group (NRG), \cite%
{NRG1,NRG2,NRG3} but as the other powerful methods it is expensive from the
numerical point of view and presents some limitations; the NRG describes
very well the Kondo Physics but presents convergence problems in the extreme
Kondo limit. \cite{Costi96} Until now it was missing a dynamical theory that
describes the transition from the weak coupling limit $U<<\Delta $ to the
strong coupling limit $U>>\Delta $, where $U$ is the Coulomb repulsion and $%
\Delta =\pi V^{2}\rho _{c}$, and $V$ is the hybridization and $\rho _{c}$ is
the density of states of the conduction electrons.

The main objective of this paper is to fill the gap described above: the
atomic approach is able to describe quantitatively all the regimes of the
SIAM in the weak, intermediate and strong correlation limits of the model.
Due to the simplicity of its implementation (practically all the method is
analytical) and very low computational cost (a density of states curve can
be obtained in few seconds or less), the atomic approach is a good candidate
to describe strongly correlated impurity systems that exhibits the Kondo
effect, like quantum dots \cite{Thiago1_2006} or carbon nanotubes. \cite%
{Thiago2_2006} The atomic approach does not substitute powerful
computational methods like  NRG,DMFT or DMRG, but can be a first choice to describe systems where the Kondo physics is relevant. In an earlier work \cite%
{Thiago06} we developed the atomic approach for the strong coupling limit ($%
U\rightarrow \infty $), and in the present work we present the atomic
approach for finite U.

The Kondo effect for both the impurity and the lattice are described by the
Anderson Hamiltonian, which considers systems that have two types of
electrons: the localized electrons ($f$-electrons), that are strongly
correlated, and the conduction electrons ($c$-electrons), that can be
described as free. The Anderson model allows the interchange between the $f$
and the $c$ electrons through the hybridization interaction, and to study
this problem we shall consider the cumulant expansion of the periodic
Anderson model (PAM) employing the hybridization as perturbation. \cite{FFM}
From the cumulant expansion we obtain formal expressions for the exact
one-electron Green's functions (GF) in terms of effective cumulants, that
are as difficult to calculate as the exact GF, and our approximation
consists in substituting these effective cumulants by those of the atomic
case of the Anderson model, that is exactly soluble. In general the Anderson
model does not have analytical solutions, but when the energies of all the $%
N $ conduction states have collapsed (the conduction band has zero width)
and the hybridization is local (i.e. independent of the wave vector $\mathbf{%
k}$), the Anderson Hamiltonian has an exact solution, and the electronic
Green's functions can be calculated analytically. In this case it is only
necessary to study a single site, that can hold up to four electrons: two $f$%
-electrons and two $c$-electrons. With the $f$-electrons we can build four
states: one $\mid 0\rangle $ with no electrons, two with one electron with
spin up$\mid +\rangle $ or with spin down$\mid -\rangle $ and one $\mid
d\rangle $ with the two electrons at the same site. In a similar way we can
build up four states $\mid 0\rangle ,\mid \uparrow \rangle ,\mid \downarrow
\rangle ,\mid \uparrow \downarrow \rangle $ with the $c$-electrons, and for
a given site we have then a vector space of dimension sixteen. The
corresponding Anderson Hamiltonian is then easily diagonalized, giving
sixteen eigenenergies and eigenstates that we shall name the atomic Anderson
solution (AAS), and this minimal Anderson system contains already the
fundamental excitations that generate the Kondo effect. To chose the
position of the conduction band of the AAS, we impose the satisfaction of
Friedel's sum rule.

In Sec. \ref{sec2} we present a brief review of the basic equations of the
atomic approach formalism developed in an earlier publication. \cite{RMP} In
Sec. \ref{sec3} we discuss in detail the equations and approximations
employed in the development of the atomic approach. In Sec. \ref{sec4} we
define the completeness and calculate the occupation numbers. In Sec. \ref%
{sec5} we discuss what criteria we shall use to determine the parameters of
the AAS: the satisfaction of the completeness or the Friedel sum rule. We
also present a discussion of the emergence of the Kondo peak from the weak
to the strong coupling regime. In Sec. \ref{sec6} we present the results of
the formation of the Kondo peak for a fixed $E_{f}$ level and variable
correlation $U$. In Sec. \ref{sec7} we fix the correlation $U$ and vary the $%
E_{f}$ level describing the empty dot, the intermediate valence (IV), the
Kondo and the magnetic regimes. To show the usefulness of the atomic
approach, we calculate in Sec. \ref{sec8} the conductance of a side coupled
QD. \cite{QDnosso,Kobayashy2004} In Sec. \ref{sec8} we present the
conclusions of the work and in the appendix \ref{ApA} we give the details of
the analytic calculation of the exact Green's function in the limit of zero
conduction bandwidth, as well as the relevant atomic Green's functions.

\section{The Anderson Hamiltonian}

\label{sec2}

In this section we present a brief review of the basic equations of the
atomic approach formalism developed in an earlier publication \cite{RMP}.
The Hamiltonian for the Anderson lattice with finite U is given by 
\begin{eqnarray}
H & =\sum_{\vec{k},\sigma}E_{\mathbf{k},\sigma}C_{\mathbf{k},\sigma
}^{\dagger}C_{\mathbf{k},\sigma}+\sum_{j,\sigma}E_{\sigma\ }f_{j,\sigma
}^{\dagger}f_{j,\sigma}  \notag \\
& +U\sum_{j}n_{j,\sigma}n_{j,\overline{\sigma}}+H_{h},  \label{Eq1}
\end{eqnarray}
where the operators $C_{\mathbf{k},\sigma }^{\dagger }$ and $C_{\mathbf{k}%
,\sigma }$ are the creation and destruction operators of the conduction band
electrons (c-electrons) with wave vector $\mathbf{k}$, component of spin $%
\sigma $ and energies $E_{\mathbf{k},\sigma }$. The $f_{i,\sigma }^{\dagger
} $ and $f_{i,\sigma }$ are the corresponding operators for the $f$%
-electrons in the Wannier localized state at site j, with site independent
energy $E_{\sigma }$ and spin component $\sigma $. The third term is the
Coulomb repulsion between the localized electrons at each site, where $%
n_{j,\sigma }=f_{j,\sigma }^{\dagger }f_{j,\sigma }$ is the number of $f$%
-electrons with spin component $\sigma $ at site $j$ and the symbol $%
\overline{\sigma }$ denotes the spin component opposite to $\sigma $. The
fourth term $H_{h}$ describes the hybridization between the localized and
conduction electrons

\begin{equation}
H_{h}=\sum_{j,\mathbf{k},\sigma }(V_{j,\mathbf{k},\sigma }f_{j,\sigma
}^{\dagger }C_{\mathbf{k},\sigma }+V_{j,\mathbf{k},\sigma }^{\ast }C_{%
\mathbf{k},\sigma }^{\dagger }f_{j,\sigma }).  \label{Eq2}
\end{equation}

The hybridization constant $V_{j,\mathbf{k},\sigma }$ in this equation is
given by 
\begin{equation}
V_{j,\mathbf{k},\sigma }=\frac{1}{\sqrt{N_{s}}}V_{\sigma }(\mathbf{k})\exp {%
(i\mathbf{k}.\mathbf{R}_{j}),}  \label{Eq2.1}
\end{equation}%
and when the Hubbard operators are introduced into Eq.~(\ref{Eq2}) , the
hybridization Hamiltonian $H_{h}$ is transformed into: 
\begin{equation}
H_{h}=\sum_{jba,\mathbf{k}\sigma }\left( V_{jba,\mathbf{k}\sigma
}X_{j,ba}^{\dagger }C_{\mathbf{k}\sigma }+V_{jba,\mathbf{k}\sigma }^{\ast
}C_{\mathbf{k}\sigma }^{\dagger }X_{j,ba}\right) ,  \label{Eq2.2}
\end{equation}%
where the label $\alpha =(b,a)$ in $H_{h}$ describes the transition $\mid
a>\rightarrow \mid b>$, and the local state $\mid a>$ has one electron more
than the state $\mid b>$. There are four local states $\mid 0>$, $\mid +>$, $%
\mid ->$ and $\mid d>=\mid +,->$ per site, and there are only four $X$
operators that destroy one local electron at a given site.

The identity relation in the reduced space of the localized states at site $%
j $ is

\begin{equation}  \label{Hub8}
X_{j , o o} + X_{j, \sigma\sigma} + X_{j , \overline{\sigma} \overline{\sigma%
}} + X_{j ,dd}= I ,
\end{equation}
where $\overline{\sigma}=-\sigma$, and the four $X_{f,aa}$ are the
projectors into the corresponding states $\mid f,a\rangle$. The occupation
numbers on the impurity $n_{f,a}=<X_{f,aa}>$ satisfy the \textquotedblleft
completeness\textquotedblright\ relation 
\begin{equation}
n_{f,0}+n_{f,\sigma}+n_{f,\overline{\sigma}}+n_{f,d}=1.  \label{completeza}
\end{equation}

We use the index $I_{x}=1,2,3,4$, defined in Table I, to characterize these $%
X$ operators:

\begin{center}
\begin{table}[tbh]
\begin{tabular}{|l|l|l|l|l|}
\hline
$I_{x}$ & $1$ & $2$ & $3$ & $4$ \\ \hline
$\alpha=(b,a)$ & $(0,+)$ & $(0,-)$ & $(-,d)$ & $(+,d)$ \\ \hline
\end{tabular}%
\caption[TABLE I]{Representation of the possible transitions present in the
finite $U$ atomic SIAM Hamiltonian. $I_{x}=1,3$ destroy one electron with
spin up and $I_{x}=2,4$ destroy one electron with spin down. We use $\protect%
\sigma=+$ and $\protect\sigma=-$ instead of $\protect\sigma=\uparrow$ and $%
\protect\sigma=\downarrow$ to emphasize that the spin belongs to a local
electron.}
\label{E5.0}
\end{table}
\end{center}

To simplify the calculation we now introduce the two matrices 
\begin{equation}
\left\{ \mathbf{M}\right\} _{_{\alpha ,\alpha ^{\prime }}}=M_{\alpha \alpha
^{\prime }}^{eff}(\mathbf{k},z,u),  \label{ApD23}
\end{equation}%
and%
\begin{equation}
\left\{ \mathbf{W}\right\} _{_{\alpha ^{\prime },\alpha }}=W_{\alpha
^{\prime },\alpha }\left( \mathbf{k},\sigma ,z\right) ,  \label{ApD24}
\end{equation}%
where $\mathbf{M}$ is the effective cumulant matrix \cite{RMP} and the
matrix elements of $\mathbf{W}$ employed in the PAM calculation are defined
by 
\begin{equation*}
W_{\alpha ^{\prime },\alpha }\left( \mathbf{k},\sigma ,z_{n}\right)
=V(\alpha ^{\prime },\mathbf{k},\sigma )V^{\ast }(\alpha ,\mathbf{k},\sigma
)\ \mathcal{G}_{c,\sigma }^{0}\left( \mathbf{k},z_{n}\right) ,
\end{equation*}%
where \ $z_{n}=i\omega _{n}$ are the Matsubara frequencies and 
\begin{equation}
\mathcal{G}_{c,\sigma }^{0}\left( \mathbf{k},z_{n}\right) =\frac{-1}{%
z_{n}-\varepsilon \left( \mathbf{k},\sigma \right) },  \label{Eq3.14}
\end{equation}%
is the free GF of the conduction electrons. A related matrix appears in the
impurity case 
\begin{equation*}
\left\{ \mathbf{W}\right\} _{_{\alpha ^{\prime },\alpha }}=W_{\alpha
^{\prime },\alpha }\left( \sigma ,z\right) .
\end{equation*}%
with matrix elements defined by 
\begin{equation*}
W_{\alpha ^{\prime },\alpha }\left( \sigma ,z\right) =\frac{1}{N_{s}}\sum_{%
\mathbf{k}}V(\alpha ^{\prime },\mathbf{k},\sigma )V^{\ast }(\alpha ,\mathbf{k%
},\sigma )\ \mathcal{G}_{c,\sigma }^{0}\left( \mathbf{k},z\right) .
\end{equation*}

The hybridization is spin independent in the Anderson model, so then

\begin{eqnarray}
V(0\sigma,\mathbf{k},\bar{\sigma}) & =V(\bar{\sigma}d,\mathbf{k},\bar{\sigma}%
)=  \notag \\
V(0\bar{\sigma},\mathbf{k},\sigma) & =V(\sigma d,\mathbf{k},\sigma)=0.
\end{eqnarray}

We shall assume a mixing that is only local, so that $V_{\sigma }(\mathbf{k}%
) $ in Eq. (\ref{Eq2.1}) \ is $\mathbf{k}$ independent, and in Eq.(\ref%
{Eq2.2}) we then have%
\begin{eqnarray}
V(0\sigma ,\mathbf{k},\sigma ) &=&V,  \label{E5.1} \\
V(\bar{\sigma}d,\mathbf{k},\sigma ) &=&\sigma V,  \label{E5.2}
\end{eqnarray}%
where we have also assumed that $V_{\sigma }(\mathbf{k})$ is independent of $%
\sigma =\pm 1$.

We shall use below the matrix $\mathbf{A=W.M}$, with matrix elements 
\begin{equation*}
A_{\alpha \alpha ^{\prime }}\left( \mathbf{k},\sigma ,z\right) \equiv \left( 
\mathbf{W\cdot M}\right) _{\alpha \alpha ^{\prime }}=
\end{equation*}%
\begin{equation}
\sum_{\alpha _{1}}W_{\alpha \alpha _{1}}\mathbf{\left( \mathbf{k},\sigma
,z\right) \ }M_{\alpha _{1}\alpha ^{\prime }}\left( \mathbf{k},\sigma
,z\right) ,  \label{Eq3.15}
\end{equation}%
and when the Hamiltonian is spin independent or commutes with the $z$
component of the spin, the $4\times 4$ matrices $\mathbf{G}^{ff}$, $\mathbf{M%
}$,~ $\mathbf{W}$ and $\mathbf{A}$ can be diagonalized into two $2\times 2$
matrices, e.g.:

\begin{equation}
\mathbf{G}^{ff}=%
\begin{pmatrix}
\mathbf{G}_{\uparrow }^{ff} & 0 \\ 
0 & \mathbf{G}_{\downarrow }^{ff}%
\end{pmatrix}%
.  \label{E5.3}
\end{equation}

In this matrix the indexes $I_{x}$ defined in Table I have been rearranged,
so that $I_{x}=1,3$ appear in $\mathbf{G}_{\uparrow}^{ff}$ and $I_{x}=2,4$
appear in $\mathbf{G}_{\downarrow}^{ff}$.

Employing Eqs. (\ref{E5.1},\ref{E5.2}) we find for the PAM

\begin{eqnarray}
\mathbf{W}_{\uparrow }\left( \mathbf{k},z\right) &=&\ \left\vert
V\right\vert ^{2}\ \mathcal{G}_{c,\uparrow }^{0}\left( \mathbf{k},z\right) \ 
\begin{pmatrix}
1 & 1 \\ 
1 & 1%
\end{pmatrix}%
,  \label{E5.4a} \\
\mathbf{W}_{\downarrow }\left( \mathbf{k},z\right) &=&\ \left\vert
V\right\vert ^{2}\ \mathcal{G}_{c,\downarrow }^{0}\left( \mathbf{k},z\right)
\ 
\begin{pmatrix}
1 & -1 \\ 
-1 & 1%
\end{pmatrix}%
,  \label{E5.4b}
\end{eqnarray}%
where $\mathcal{G}_{c,\sigma }^{0}\left( \mathbf{k},z\right) $ is given by
Eq.(\ref{Eq3.14}). For an impurity located at the origin we find instead 
\begin{eqnarray}
\mathbf{W}_{\uparrow }\left( z\right) &=&\left\vert V\right\vert ^{2}\varphi
_{\uparrow }(z)%
\begin{pmatrix}
1 & 1 \\ 
1 & 1%
\end{pmatrix}%
,  \label{E5.5a} \\
\mathbf{W}_{\downarrow }\left( z\right) &=&\left\vert V\right\vert
^{2}\varphi _{\downarrow }(z)\ 
\begin{pmatrix}
1 & -1 \\ 
-1 & 1%
\end{pmatrix}%
,  \label{E5.5b}
\end{eqnarray}%
where%
\begin{equation}
\varphi _{\sigma }(z)=\frac{1}{N_{s}}\sum_{\mathbf{k}}\ \mathcal{G}%
_{c,\sigma }^{0}\left( \mathbf{k},z\right) .  \label{E5.6a}
\end{equation}%
For a rectangular band with half-width $D$ in the interval $[A,B],$ with $%
B=A+2D$ we then find 
\begin{equation}
\varphi _{\sigma }(z)=\frac{1}{2D}\ \ln \left( \frac{z-B+\mu }{z+A+\mu }%
\right) ,  \label{E5.6}
\end{equation}%
where the chemical potential $\mu $ appears in $\varphi _{\sigma }(z)$
because of the $\varepsilon \left( \mathbf{k},\sigma \right) =E_{\mathbf{%
k,\sigma }}-\mu $ in $\mathcal{G}_{c,\sigma }^{0}\left( \mathbf{k},z\right) $%
..

From our earlier work \cite{RMP} we obtain the exact formal Green's
functions, both for the PAM and for the SIAM: 
\begin{equation}
\mathbf{G}_{\sigma }^{ff}=\mathbf{M}_{\sigma }\mathbf{\cdot }\left( \mathbf{%
I-A}_{\sigma }\right) ^{-1},  \label{E5.7}
\end{equation}%
and from this equation follows 
\begin{equation}
\mathbf{M}_{\sigma }\mathbf{=}\left( \mathbf{I+G}_{\sigma }^{ff}\cdot 
\mathbf{W}_{\sigma }\right) ^{-1}\cdot \mathbf{G}_{\sigma }^{ff}.
\label{E5.8}
\end{equation}

\section{The atomic approach for the single impurity Anderson model (SIAM)}

\label{sec3}

Defining the exact cumulants as

\begin{equation}
\mathbf{M}_{\uparrow }=%
\begin{pmatrix}
m_{11} & m_{13} \\ 
m_{31} & m_{33}%
\end{pmatrix}%
\hspace{20pt}\mathrm{;\hspace{20pt}}\mathbf{M}_{\downarrow }=%
\begin{pmatrix}
m_{22} & m_{24} \\ 
m_{42} & m_{44}%
\end{pmatrix}%
,  \label{E5.97}
\end{equation}%
one obtains the exact Green's functions $\mathbf{G}_{\sigma }^{ff}(i\omega )$
of the localized $f$ electrons by performing the matrix inversion in Eq. (%
\ref{E5.7}), employing Eqs. (\ref{E5.5a},\ref{E5.5b}) and Eq. (\ref{E5.97}): 
\begin{widetext}
\begin{equation}
\mathbf{G}_{\uparrow}^{ff}(i\omega)=\begin{pmatrix}
G^{ff}_{11} & G^{ff}_{13} \\
G^{ff}_{31} & G^{ff}_{33}%
\end{pmatrix}%
=\frac{%
\begin{pmatrix}
m_{11} & m_{13} \\
m_{31} & _{m33}%
\end{pmatrix}
-\left\vert V\right\vert ^{2}\varphi_{\uparrow}((i\omega))\left(
m_{11}m_{33}-m_{13}m_{31}\right)
\begin{pmatrix}
1 & -1 \\
-1 & 1%
\end{pmatrix}
}{1-\left\vert V\right\vert ^{2}\varphi_{\uparrow}((i\omega))\left(
m_{11}+m_{33}+m_{13}+m_{31}\right) } ,  \label{E5.12}
\end{equation}
\begin{equation}
\mathbf{G}_{\downarrow}^{ff}(i\omega)=\begin{pmatrix}
G^{ff}_{22} & G^{ff}_{24} \\
G^{ff}_{42} & G^{ff}_{44}%
\end{pmatrix}%
=
\frac{%
\begin{pmatrix}
m_{22} & m_{24} \\
m_{42} & _{m44}%
\end{pmatrix}
-\left\vert V\right\vert ^{2}\varphi_{\downarrow}((i\omega))\left(
m_{22}m_{44}-m_{24}m_{42}\right)
\begin{pmatrix}
1 & 1 \\
1 & 1%
\end{pmatrix}
}{1-\left\vert V\right\vert ^{2}\varphi_{\downarrow}((i\omega))\left(
m_{22}+m_{44}-m_{24}-m_{42}\right) }.  \label{E5.13}
\end{equation}

In the same way we can obtain the conduction $\mathbf{G}_{\sigma}^{cc}(\mathbf{k},\mathbf{k}^{\prime},i\omega)$
and the cross $\mathbf{G}_{\sigma}^{cf}(\mathbf{k},i\omega)$  Green's functions; a detailed derivation can be found in  \cite{RMP}

$$
\mathbf{G}_{\uparrow}^{cc}(\mathbf{k},\mathbf{k}^{\prime},i\omega
)=\mathcal{G}_{c,\uparrow}^{0}\left(  \mathbf{k},i\omega\right)  \delta\left(
\mathbf{k,k}^{\prime}\right)  + \nonumber
$$
\begin{equation}
\frac{\left\vert V\right\vert ^{2}}{N_{s}%
}\mathcal{G}_{c,\uparrow}^{0}\left(  \mathbf{k},i\omega\right)  \frac{\left(
m_{11}+m_{33}+m_{13}+m_{31}\right)  }{1-\left\vert V\right\vert ^{2}%
\varphi_{\uparrow}(i\omega)\left(  m_{11}+m_{33}+m_{13}+m_{31}\right)
}\mathcal{G}_{c,\uparrow}^{0}\left(  \mathbf{k}^{\prime},i\omega\right) ,
\label{E5.21b}%
\end{equation}

$$
\mathbf{G}_{\downarrow}^{cc}(\mathbf{k},\mathbf{k}^{\prime},i\omega
)=\mathcal{G}_{c,\downarrow}^{0}\left(  \mathbf{k},i\omega\right)
\delta\left(  \mathbf{k,k}^{\prime}\right)  + \nonumber
$$
\begin{equation}
\frac{\left\vert V\right\vert
^{2}}{N_{s}}\mathcal{G}_{c,\downarrow}^{0}\left(  \mathbf{k},i\omega\right)
\frac{\left(  mm_{22}+m_{44}-m_{24}-m_{42}\right)  }{1-\left\vert V\right\vert
^{2}\varphi_{\downarrow}(i\omega)\left(  m_{22}+m_{44}-m_{24}-m_{42}\right)
}\mathcal{G}_{c,\downarrow}^{0}\left(  \mathbf{k}^{\prime},i\omega\right) ,
\label{E5.21c}%
\end{equation}
and  the cross Green function $G_{\sigma}^{cf}$ is defined by a column vector with two components as \cite{RMP}
\begin{equation}
G_{\sigma}^{cf}=
\begin{pmatrix}
G_{0\sigma,\sigma}^{cf}\\
G_{\sigma d,\sigma}^{cf}%
\end{pmatrix} ,
\end{equation}
\begin{equation}
\mathbf{G}_{\uparrow}^{cf}(\mathbf{k},i\omega)=-\frac{V}{\sqrt{N_{s}}%
}\mathcal{G}_{c,\uparrow}^{0}\left(  \mathbf{k},i\omega\right)  \frac{%
\begin{pmatrix}
m_{11}+m_{31} & ,m_{13}+m_{33}%
\end{pmatrix}
}{1-\left\vert V\right\vert ^{2}\varphi_{\uparrow}(i\omega)\left(
m_{11}+m_{33}+m_{13}+m_{31}\right)  }\label{E5.19b} ,
\end{equation}%
\begin{equation}
\mathbf{G}_{\downarrow}^{cf}(\mathbf{k},i\omega)=-\frac{V}{\sqrt{N_{s}}%
}\mathcal{G}_{c,\downarrow}^{0}\left(  \mathbf{k},i\omega\right)  \frac{%
\begin{pmatrix}
m_{22}-m_{42} & ,m_{24}-m_{44}%
\end{pmatrix}
}{1-\left\vert V\right\vert ^{2}\varphi_{\downarrow}(i\omega)\left(
m_{22}+m_{44}-m_{24}-m_{42}\right)  }\label{E5.19c} .
\end{equation}
\end{widetext}

The calculation of the exact effective cumulants $M_{\sigma }$ is as
difficult as that of the exact $\mathbf{G}_{\sigma }^{ff}$, and the atomic
approach consists in using instead the effective cumulants of a similar
model that is exactly soluble. The atomic limit of the SIAM is just a
particular case of the general model, and we shall then use the AAS to
calculate the exact Green's function $G_{\sigma }^{ff,at}(z)$ of the atomic
problem, which then satisfies a relation of the same form of Eq.~(\ref{E5.7}%
): 
\begin{equation}
\mathbf{G}_{\sigma }^{ff,at}=\mathbf{M}_{\sigma }^{at}\mathbf{\cdot }\left( 
\mathbf{I-W_{\sigma }^{o}M_{\sigma }^{at}}\right) ^{-1}.  \label{E5.7a}
\end{equation}

From this equation we obtain the exact atomic cumulant $\mathbf{M}_{\sigma
}^{at}$ 
\begin{equation}
\mathbf{M}_{\sigma }^{at}\mathbf{=}\left( \mathbf{I+G}_{\sigma
}^{ff,at}\cdot \mathbf{W}_{\sigma }^{o}\right) ^{-1}\cdot \mathbf{G}_{\sigma
}^{ff,at},  \label{E5.8a}
\end{equation}%
which satisfies Eq. (\ref{E5.8}). For an impurity located at the origin we
change Eqs.~(\ref{E5.5a},\ref{E5.5b}) into 
\begin{align}
\mathbf{W}_{\uparrow }^{o}\left( z\right) & =\left\vert \Delta \right\vert
^{2}\varphi _{\uparrow }^{o}(z)%
\begin{pmatrix}
1 & 1 \\ 
1 & 1%
\end{pmatrix}%
,  \label{E5.55a} \\
\mathbf{W}_{\downarrow }^{o}\left( z\right) & =\left\vert \Delta \right\vert
^{2}\varphi _{\downarrow }^{o}(z)\ 
\begin{pmatrix}
1 & -1 \\ 
-1 & 1%
\end{pmatrix}%
,  \label{E5.55b}
\end{align}%
where 
\begin{equation}
\varphi _{\sigma }^{o}(z)=\frac{-1}{z-\varepsilon _{o}-\mu },
\label{Eq3.144}
\end{equation}%
is obtained by replacing all the $\mathcal{G}_{c,\sigma }^{0}\left( \mathbf{k%
},z\right) $ in Eq. (\ref{E5.6a}) by those corresponding to the zeroth-width
band located at $\varepsilon _{0}$, namely the bare conduction Green
function. This procedure overestimates the contribution of the $c$
electrons, \cite{Alascio79} because we concentrate them at a single energy
level $\varepsilon _{o}$, and to moderate this effect we replace $V^{2}$ by $%
\Delta ^{2}$ in Eqs. (\ref{E5.55a}-\ref{E5.55b}), where $\Delta =\pi
V^{2}/2D $ is the Anderson parameter.

The atomic approach consists in substituting the exact effective cumulant $%
M_{\sigma }$, that appears in Eqs. (\ref{E5.12}-\ref{E5.19c}), by the exact
atomic one $M_{\sigma }^{at}$, which is defined by Eqs. (\ref{E5.8a}-\ref%
{Eq3.144}). We call $M_{\sigma }^{ap}$ this approximate cumulant, and we
make the substitution

\begin{equation}
M_{\sigma }\rightarrow M_{\sigma }^{ap}  \label{M34}
\end{equation}%
Performing the matrix inversion in Eq. (\ref{E5.8a}) and employing Eqs. (\ref%
{E5.55a},\ref{E5.55b}) it is now straightforward to obtain 
\begin{widetext}
\begin{equation}
\mathbf{M}_{\uparrow}^{ap}(i\omega)=
\begin{pmatrix}
m^{ap}_{11} & m^{ap}_{13} \\
m^{ap}_{31} & m^{ap}_{33}%
\end{pmatrix}%
=\frac{%
\begin{pmatrix}
g_{11} & g_{13} \\
g_{31} & g_{33}
\end{pmatrix}
+\left\vert \Delta\right\vert ^{2}\varphi_{\uparrow}^{o}(i\omega)\left(
g_{11} g_{33} - g_{13} g_{31}\right)
\begin{pmatrix}
1 & -1 \\
-1 & 1%
\end{pmatrix}
}{1+\left\vert \Delta \right\vert ^{2}\varphi_{\uparrow}^{o}(i\omega)\left(
g_{11} + g_{33} + g_{13} + g_{31}\right) } , \label{5.121}
\end{equation}
\begin{equation}
\mathbf{M}_{\downarrow}^{ap}(i\omega)=\begin{pmatrix}
m^{ap}_{22} & m^{ap}_{24} \\
m^{ap}_{42} & m^{ap}_{44}%
\end{pmatrix}%
=\frac{%
\begin{pmatrix}
g_{22} & g_{24} \\
g_{42} & g_{44}
\end{pmatrix}
+\left\vert \Delta \right\vert ^{2}\varphi_{\downarrow}^{o}(i\omega)\left(
g_{22} g_{44} - g_{24} g_{42}\right)
\begin{pmatrix}
1 & 1 \\
1 & 1%
\end{pmatrix}
}{1+\left\vert \Delta \right\vert ^{2}\varphi_{\downarrow}^{o}(i\omega)\left(
g_{22} + g_{44} - g_{24}- g_{42}\right) } , \label{5.122}
\end{equation}
\end{widetext}where the $g_{ij}$ are the atomic Green's functions of the $f$
electrons that are calculated in Appendix \ref{ApA}. Substituting now the
approximate $M_{\sigma }^{ap}$'s in Eqs.~(\ref{E5.12}-\ref{E5.13}) we obtain
the Green's functions of the localized $f$ electrons in the atomic approach.

We should stress that it is essential to use $\varphi _{\sigma }^{o}(z)$
rather than $\varphi _{\sigma }(z)$ in Eqs. (\ref{E5.55a},\ref{E5.55b}) to
obtain a well defined Kondo peak structure at the chemical potential $\mu $.
If we perform instead the calculation employing $\varphi _{\sigma }(z)$ we
always obtain a wrong structure with two or more peaks around the chemical
potential, as obtained in early works using the atomic solution of the
Anderson model \cite{Acirete88,Marinaro91}. In the computational calculation
we fixed the chemical potential at $\mu =0$ and varied the conduction atomic
level $\varepsilon _{0}$ in such a way that the Friedel sum rule should be
satisfied. This point will be discussed in more detail in the next section.

\section{The completeness problem and occupation numbers}

\label{sec4}

Employing Eq. (\ref{E5.3}) we separate the different quantities in two
different types: those associated with $I_{x}=1,3$ corresponds spin up
electrons and those associated with $I_{x}=2,4$ corresponds to spin down
electrons.

In the finite $U$ case, the identity operator in the space of the impurity
local $f$ states is given by 
\begin{equation}
X_{oo}+X_{\sigma \sigma }+X_{\overline{\sigma }\overline{\sigma }}+X_{dd}=I.
\label{Hub8a}
\end{equation}%
The completeness is associated to the average of this equation: 
\begin{equation}
\left\langle X_{oo}+X_{\sigma \sigma }+X_{\overline{\sigma }\overline{\sigma 
}}+X_{dd}\right\rangle =1,  \label{Occ1}
\end{equation}%
where the first term is the vacuum occupation number, the second and the
third terms are the occupation of the spin up and down respectively, and the
last term is the double occupation. Employing the notation in Table I to
identify the $X_{ba}$ operators we could then write this equation in the
form 
\begin{equation}
\left\langle X_{1}X_{1}^{\dagger }\right\rangle +\left\langle X_{1}^{\dagger
}X_{1}\right\rangle +\left\langle X_{3}X_{3}^{\dagger }\right\rangle
+\left\langle X_{3}^{\dagger }X_{3}\right\rangle =1,  \label{D04}
\end{equation}%
and all the different averages could be calculated employing the Green's
functions $G_{11}^{ff}(\omega )$ and $G_{33}^{ff}(\omega )$ defined in Eqs. (%
\ref{E5.12}) and associated with the processes $I_{x}=1,3$: 
\begin{equation}
<X_{o,o}>=<X_{1}X_{1}^{\dagger }>=\left( \frac{-1}{\pi }\right)
\int_{-\infty }^{\infty }d\omega Im(G_{11}^{ff})(1-n_{F}),  \label{G00}
\end{equation}%
\begin{equation}
<X_{+,+}>=<X_{1}^{\dagger }X_{1}>=\left( \frac{-1}{\pi }\right)
\int_{-\infty }^{\infty }d\omega Im(G_{1,1}^{ff})n_{F},  \label{Gup}
\end{equation}%
\begin{equation}
<X_{-,-}>=<X_{3}X_{3}^{\dagger }>=\left( \frac{-1}{\pi }\right)
\int_{-\infty }^{\infty }d\omega Im(G_{33}^{ff})(1-n_{F}),  \label{Gdown}
\end{equation}%
\begin{equation}
<X_{d,d}>=<X_{3}^{\dagger }X_{3}>=\left( \frac{-1}{\pi }\right)
\int_{-\infty }^{\infty }d\omega Im(G_{33}^{ff})n_{F},  \label{Gdd}
\end{equation}%
where $n_{F}(x)=1/\left[ 1+\exp (\beta x)\right] ${\ is the Fermi-Dirac
distribution. }

In a similar way we could employ the Green's functions $G_{22}^{ff}$ and $%
G_{44}^{ff}$ defined in Eqs. (\ref{E5.13}), and associated with the
processes $I_{x}=2,4$: 
\begin{equation}
\left\langle X_{2}X_{2}^{\dagger }\right\rangle +\left\langle X_{2}^{\dagger
}X_{2}\right\rangle +\left\langle X_{4}X_{4}^{\dagger }\right\rangle
+\left\langle X_{4}^{\dagger }X_{4}\right\rangle =1,  \label{D05}
\end{equation}%
\begin{equation}
<X_{o,o}>=<X_{2}X_{2}^{\dagger }>=\left( \frac{-1}{\pi }\right)
\int_{-\infty }^{\infty }d\omega Im(G_{22}^{ff})(1-n_{F}),  \label{G00L}
\end{equation}%
\begin{equation}
<X_{-,-}>=<X_{2}^{\dagger }X_{2}>=\left( \frac{-1}{\pi }\right)
\int_{-\infty }^{\infty }d\omega Im(G_{22}^{ff})n_{F},  \label{GupL}
\end{equation}%
\begin{equation}
<X_{+,+}>=<X_{4}X_{4}^{\dagger }>=\left( \frac{-1}{\pi }\right)
\int_{-\infty }^{\infty }d\omega Im(G_{44}^{ff})(1-n_{F}),  \label{GdownL}
\end{equation}%
\begin{equation}
<X_{d,d}>=<X_{4}^{\dagger }X_{4}>=\left( \frac{-1}{\pi }\right)
\int_{-\infty }^{\infty }d\omega Im(G_{44}^{ff})n_{F}.  \label{GddL}
\end{equation}

\section{The Friedel sum rule: a criteria to be satisfied}

\label{sec5}

The Friedel's sum rule (FSR) \cite{Langreth66} gives at $T=0$, a
relationship between the extra states induced below the Fermi level by a
scattering center and the phase shift at the chemical potential $\eta
_{\sigma }(\mu )$, obtained by the transference matrix $T_{ff,\sigma
}(z)=V^{2}G_{ff,\sigma }^{imp}(z)$, where $V$ is the scattering
potential. For the SIAM the extra states induced are given by the occupation
number $n_{f,\sigma }$ of the localized state, and the scattering potential
is the hybridization that affects the conduction electrons. The Friedel's
sum rule (FSR) for the Anderson impurity model can be written as \cite%
{Kang2001} 
\begin{equation}
\rho _{f\sigma }(\mu )=\frac{sin^{2}\left( \pi n_{f\sigma }\right) }{\Delta
\pi },  \label{fried}
\end{equation}%
\noindent where $\rho _{f,\sigma }(\mu )$ is the density of states of the
localized level at the chemical potential.

The atomic approach consists in substituting the exact effective cumulant $%
M_{\sigma }$, that appears in Eqs. (\ref{E5.12}-\ref{E5.19c}), by the exact
atomic one $M_{\sigma }^{at}$, which is defined by Eqs. (\ref{E5.8a}-\ref%
{Eq3.144}). The difference between the exact and the approximate GF's is
that different energies $\varepsilon _{k}$ appear in the c-electron
propagators of the effective cumulant $M_{\sigma }^{eff}(z)$, while these
energies are all equal to the atomic conduction level $\varepsilon _{0}$ in $%
M_{\sigma }^{at}(z)$. Although $M_{\sigma }^{at}(z)$ is for that reason only
an approximation, it contains all the cumulant diagrams that should be
present, and one would expect that the corresponding GF would have fairly
realistic features. One still has to decide what value of $\varepsilon _{0}$
should be taken. As the most important region of the conduction electrons is
the chemical potential $\mu $, we could choose $\varepsilon _{0}=\mu $, but
we shall consider instead that the position of the atomic conduction level
is given by $\xi =\mu \pm \delta \varepsilon _{0}$. This leaves the freedom
of choosing $\varepsilon _{0}$ so that the Friedel sum rule given by Eq. (%
\ref{fried}) should be satisfied.

In our earlier paper \cite{Thiago06}, where we developed the atomic approach
of the Anderson impurity model for infinite $U$, we imposed the satisfaction
of the completeness relation $X_{j,oo}+X_{j,\sigma \sigma }+X_{j,\overline{%
\sigma }\overline{\sigma }}=I$ instead of the fulfillment of the Friedel sum
rule. To be rigorous the FSR is only valid at temperature $T=0$, but we can
use it as an approximation at temperatures of order of the Kondo temperature 
$T_{K}$. The validity of the completeness condition is much more general
than the FSR, because completeness is valid for the whole range of
temperatures and parameters of the model. In the case of the finite $U$
Anderson model the completeness is always satisfied and we cannot use it to
determine $\varepsilon _{0}$: we shall then use the fulfillment of the
Friedel sum rule as a condition to obtain the adequate physical solution.

Employing the results of the Section \ref{sec2} and Appendix \ref{ApA} we
can calculate the ($f,c$) components of the matrix Green's functions for
finite $U$ in the atomic approach, and the corresponding spectral densities
are given by 
\begin{equation}
\rho_{f,c}(\omega )=\left( \frac{-1}{\pi }\right)
Im[G^{ff,cc}(\omega )].  \label{E6.1}
\end{equation}

\begin{figure}[th]
{\includegraphics[clip,width=0.40\textwidth,angle=-90.]{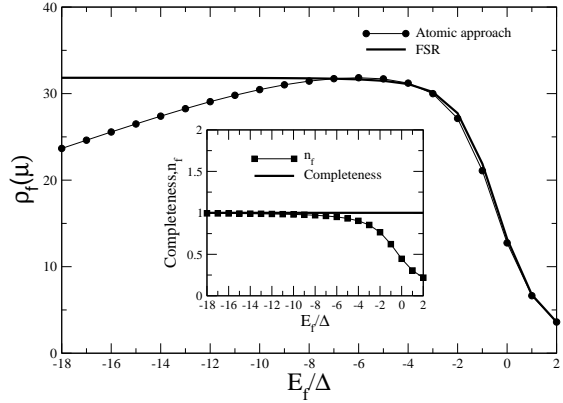}}
\caption{Density of states at the chemical potential $\protect\mu $ as
function of $E_{f}$, for $U\rightarrow \infty $ and $T=0.001\Delta $; we
impose the satisfaction of the completeness relation. The $\protect\rho %
_{f}\left( \protect\mu \right) $ in FSR was calculated with Eq. (\protect\ref%
{fried}). }
\label{Completeza1}
\end{figure}
\begin{figure}[th]
{\includegraphics[clip,width=0.40\textwidth,angle=-90.]{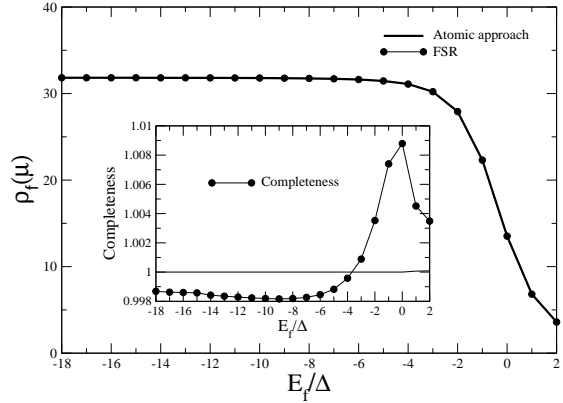}}
\caption{Density of states at the chemical potential $\protect\mu $ as a
function of $E_{f}$ for $U\rightarrow \infty $ and $T=0.001\Delta $. The $%
\protect\rho _{f}\left( \protect\mu \right) $ in FSR was calculated with Eq.
(\protect\ref{fried}). In this case we impose the satisfaction of
Friedel'sum rule.}
\label{Completeza2}
\end{figure}
\begin{figure}[th]
\includegraphics[clip,width=0.40%
\textwidth,angle=-90]{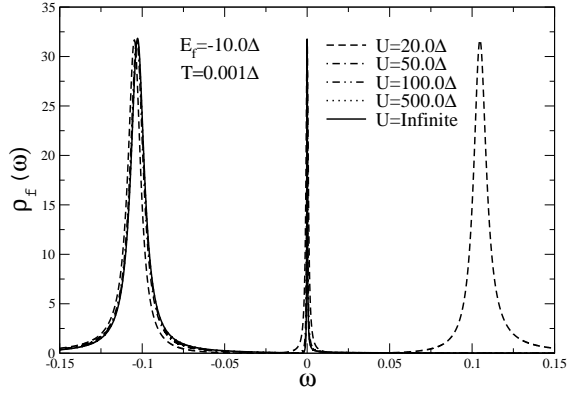}
\caption{Density of states for $T=0.001\Delta $, $E_{f}=-10.0\Delta $ and
several $U$ values: $U=20.0\Delta $, $U=50.0\Delta $, $U=100.0\Delta $, $%
U=500.0\Delta $ and $U=\infty $.}
\label{Ulimit}
\end{figure}
\begin{figure}[th]
\includegraphics[clip,width=0.40%
\textwidth,angle=-90]{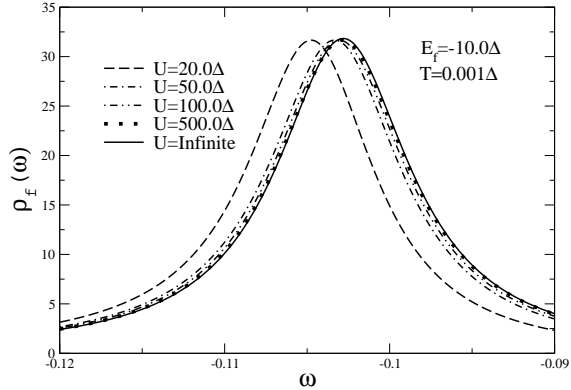}
\caption{Detail of the lower band in the density of states of the Fig. 
\protect\ref{Ulimit}.}
\label{Ulimit_Ef}
\end{figure}
\begin{figure}[th]
\includegraphics[clip,width=0.40%
\textwidth,angle=-90]{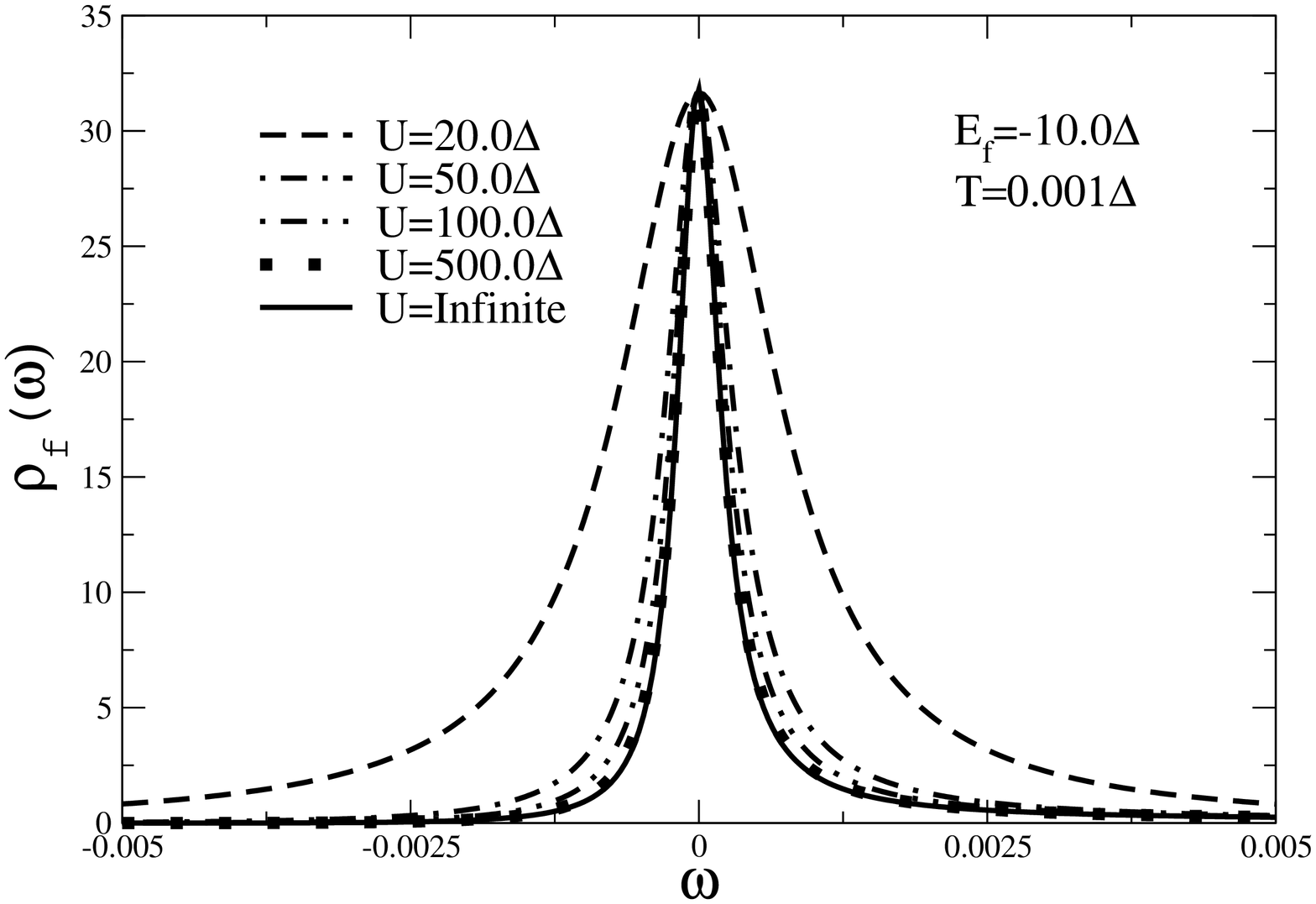}
\caption{Detail of the Kondo peak in the density of states of the Fig. 
\protect\ref{Ulimit}.}
\label{Ulimit_Kondo}
\end{figure}
In all the figures of the paper we employ $\Delta $ units, with $\Delta =\pi
V^{2}/2D=0.01$, with $D=1.0$.

In Fig. \ref{Completeza1} we calculate the density of states at the chemical
potential $\mu $ as a function of $E_{f}$, for $U\rightarrow \infty $ and $%
T=0.001\Delta $. In this case we impose the satisfaction of the completeness
and the figure shows that in the extreme Kondo region $n_{f}\simeq 1$, there
is a considerable departure of the $\rho _{f}\left( \mu \right) $ that
satisfies the Friedel sum rule; in the inset of that figure we show the
evolution of the occupation $n_{f}$ toward the Kondo limit and that the
completeness is satisfied.

In Fig. \ref{Completeza2} we impose the satisfaction of Friedel'sum rule.
According to that figure the $f$ density of states at the chemical potential 
$\mu $, satisfies the FSR. In the inset of this figure we show that the
completeness is lost but in the Kondo region the departure from completeness
is very low; the error is less than the $1\%$, which justifies the use of
FSR to determine $\varepsilon _{0}$. We shall then employ this criteria both
for $U\rightarrow \infty $ and for finite $U$, because the temperatures
involved in this effect are generally very low.

Next we study the evolution of the Kondo peak for finite $U$ when the
correlation energy $U$ increases toward $U\rightarrow \infty $. In Fig. \ref%
{Ulimit} we plot the density of states corresponding the evolution of the
Kondo peak for $T=0.001\Delta $, $E_{f}=-10.0\Delta $ and several $U$
values: $U=20.0\Delta $, $U=50.0\Delta $, $U=100.0\Delta $, $U=500.0\Delta $
and $U=\infty $. We show the upper band only for $U=20.0\Delta $. In Fig. %
\ref{Ulimit_Ef} we plot in detail the resonant band located around $E_{f}$:
the curve with $U=500.0\Delta $ is practically coincident with the one with $%
U\rightarrow \infty $. In Fig. \ref{Ulimit_Kondo} we plot in detail the
Kondo peak for the same values, and again the plot with $U=500.0\Delta $
practically coincides with the one for $U\rightarrow \infty $.

\section{The emergence of the Kondo peak}

\label{sec6}

\begin{figure}[th]
\includegraphics[clip,width=0.40\textwidth,angle=-90.0]{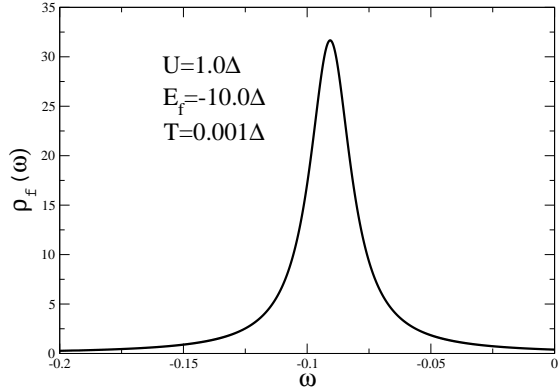}
\caption{Density of states $\protect\rho_{f}(\protect\omega)$ for $%
U=1.0\Delta$, $E_{f}=-10.0 \Delta$ and $T=0.001 \Delta$.}
\label{Fig1}
\end{figure}
\begin{figure}[th]
\includegraphics[clip,width=0.40\textwidth,angle=-90.0]{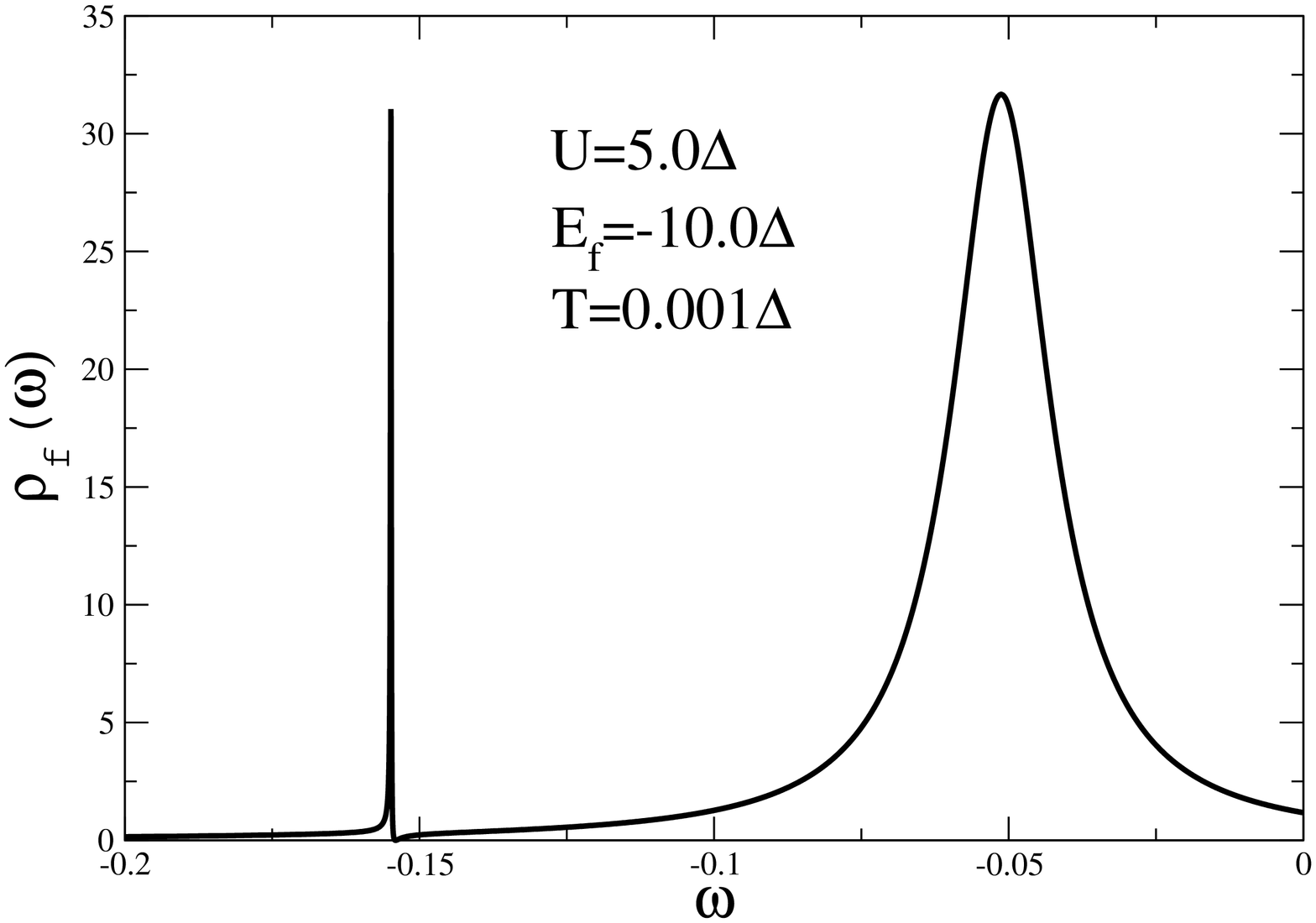}
\caption{Density of states $\protect\rho_{f}(\protect\omega)$ for $%
U=5.0\Delta$, $E_{f}=-10.0 \Delta$ and $T=0.001 \Delta$.}
\label{Fig2}
\end{figure}
\begin{figure}[th]
\includegraphics[clip,width=0.40\textwidth,angle=-90.0]{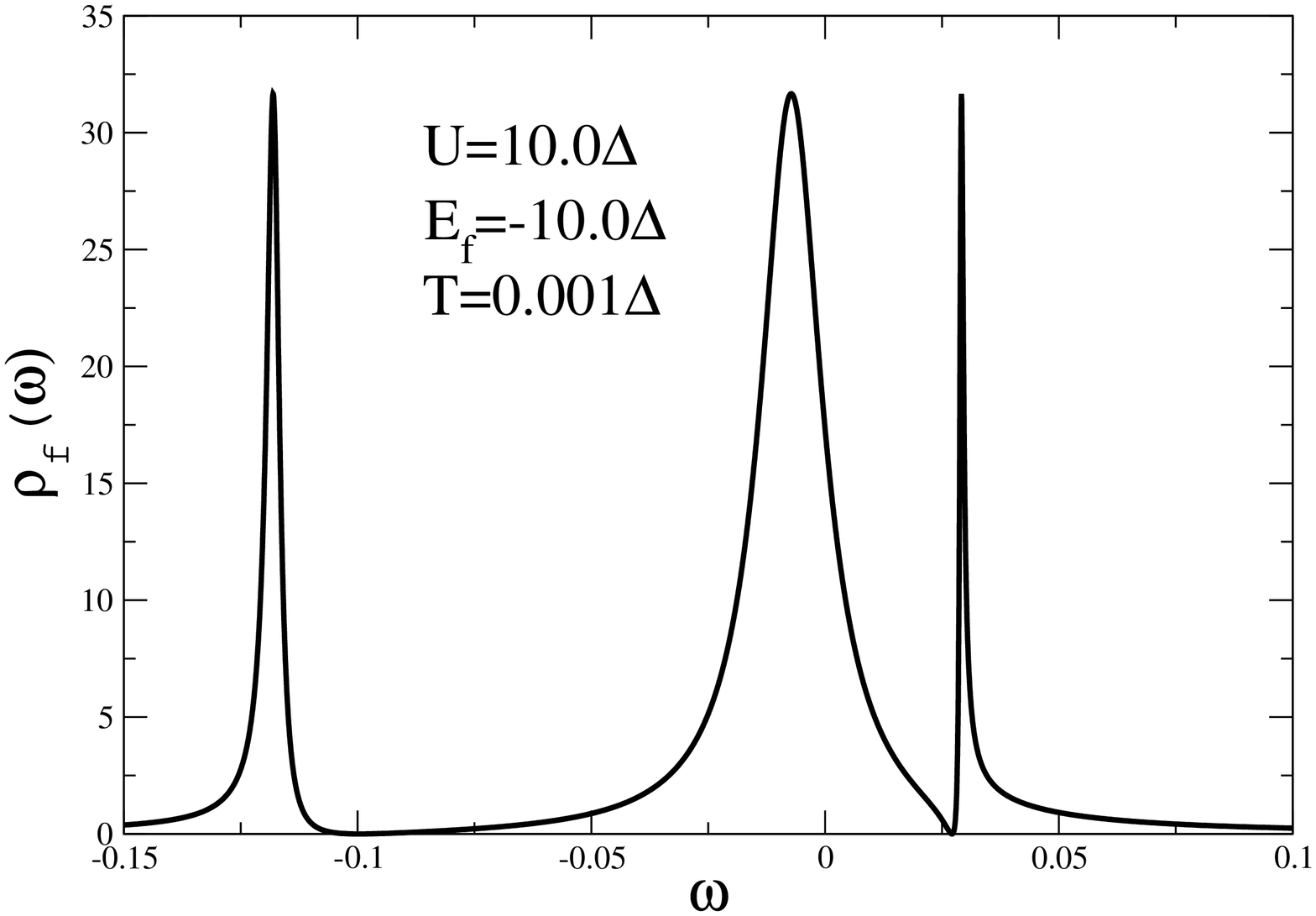}
\caption{Density of states $\protect\rho_{f}(\protect\omega)$ for $%
U=10.0\Delta$, $E_{f}=-10.0 \Delta$ and $T=0.001 \Delta$.}
\label{Fig3}
\end{figure}
\begin{figure}[th]
\includegraphics[clip,width=0.40\textwidth,angle=-90.0]{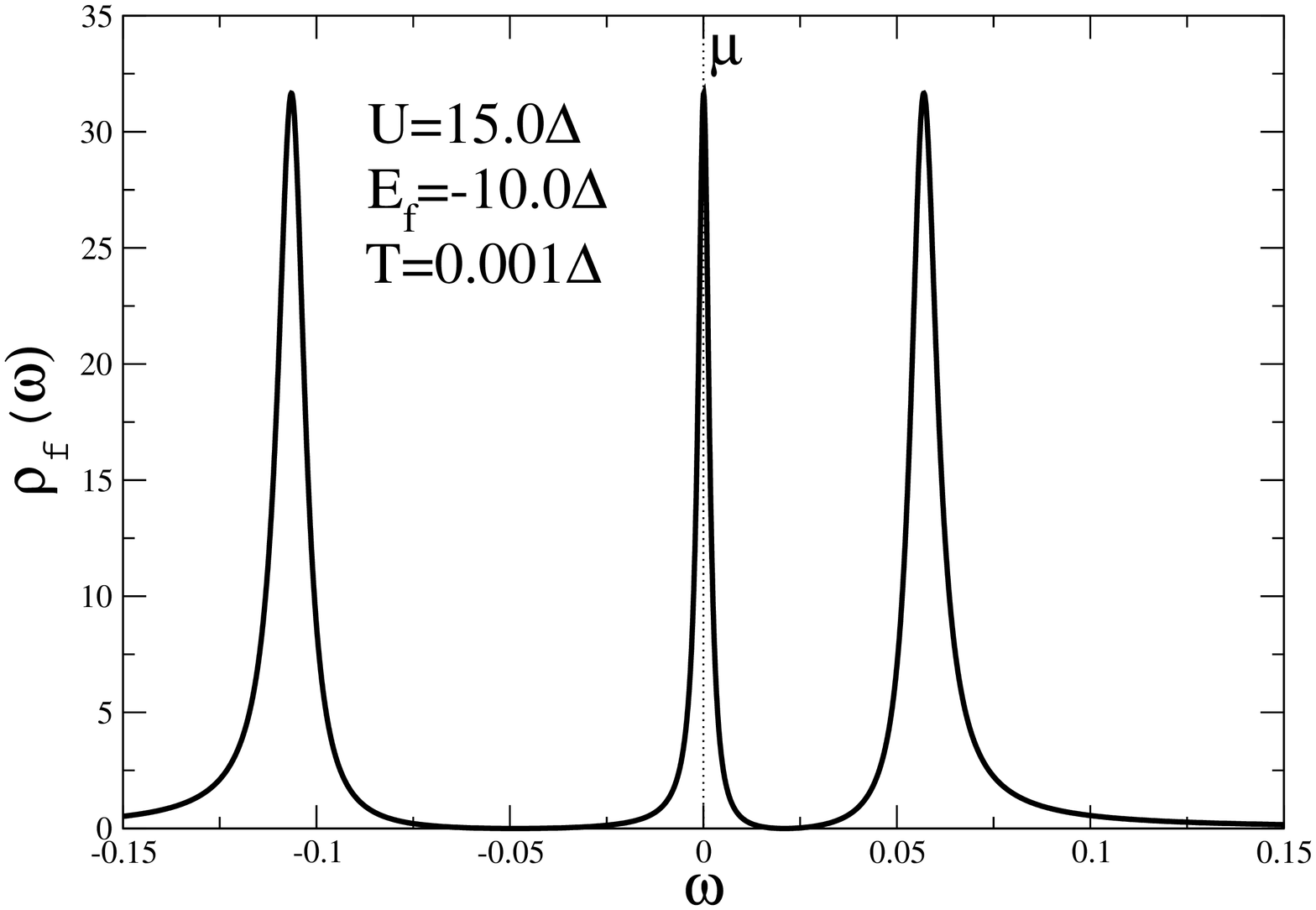}
\caption{Density of states $\protect\rho_{f}(\protect\omega)$ for $%
U=15.0\Delta$, $E_{f}=-10.0 \Delta$ and $T=0.001 \Delta$.}
\label{Fig33}
\end{figure}

\begin{figure}[th]
\includegraphics[clip,width=0.40%
\textwidth,angle=-90.0]{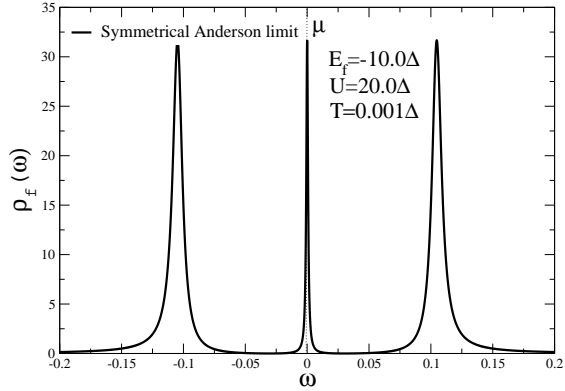}
\caption{Density of states for $T=0.001 \Delta$ , $U=20\Delta$ and $%
E_{f}=-10.0 \Delta$. This parameter set corresponds to the symmetric case of
the model and the total localized occupation number is exactly the unity $%
n_{f}=1.0$}
\label{Fig13}
\end{figure}
\begin{figure}[th]
\includegraphics[clip,width=0.40\textwidth,angle=-90.0]{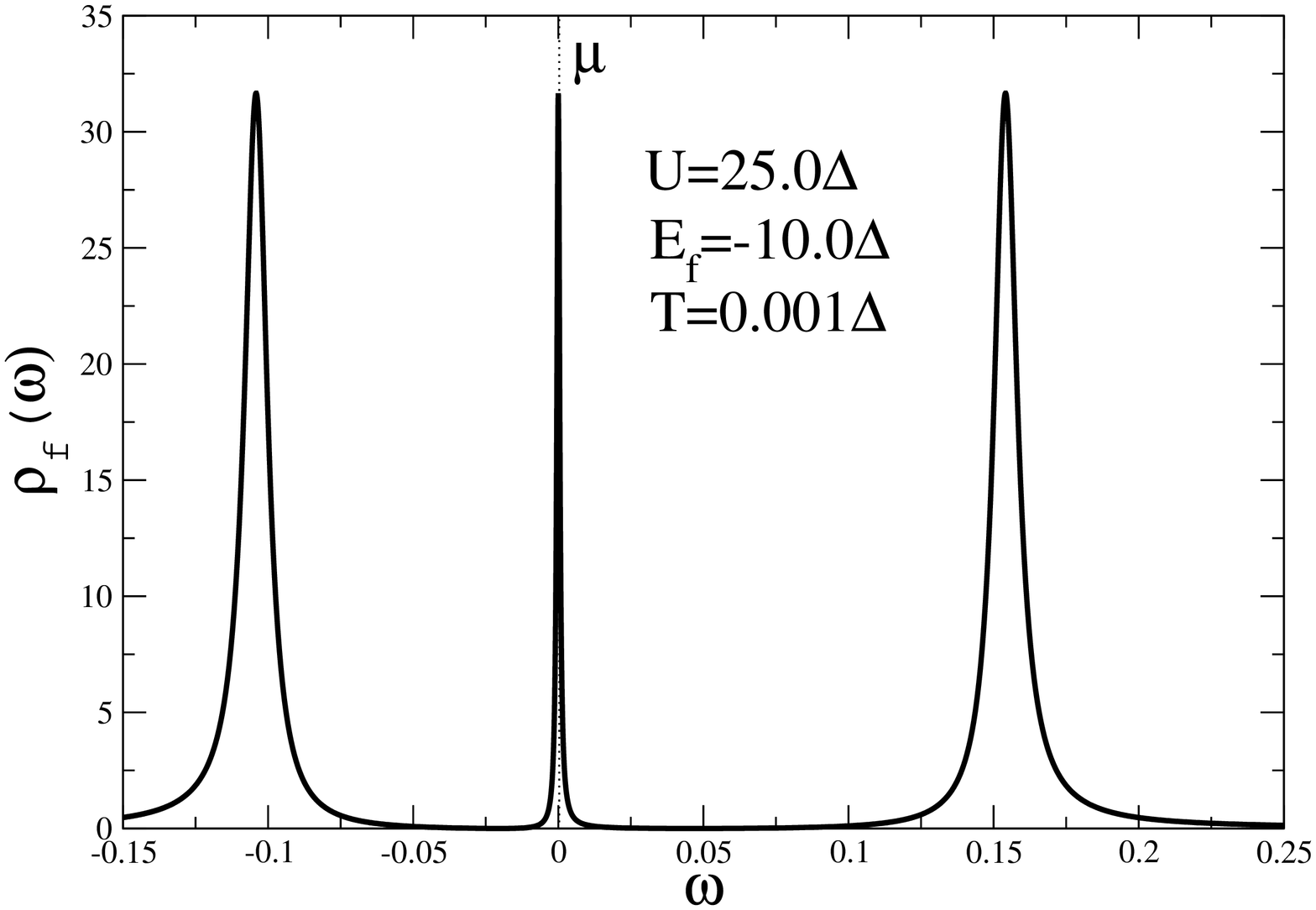}
\caption{Density of states $\protect\rho_{f}(\protect\omega)$ for $%
U=25.0\Delta$, $E_{f}=-10.0 \Delta$ and $T=0.001 \Delta$.}
\label{Fig333}
\end{figure}

In the set of Figs. (\ref{Fig1}-\ref{Fig333}) we show the evolution of the $%
f $ density of states as a function of the Coulomb repulsion $U$. We begin
with the uncorrelated limit of the finite $U$ SIAM (weak coupling limit),
and we consider $U=1.0\Delta$, $U=5.0\Delta$, $U=10.0 \Delta$, $U=15.0
\Delta $, $U=20.0 \Delta$ (this value corresponds to symmetric case) and $%
U=25.0 \Delta$.

We observe that in the Figs. (\ref{Fig1}-\ref{Fig3}), where the correlation
is weak, the three peak structure characteristic of the SIAM starts to
appear, but in this region the Kondo peak is not yet formed. However, for $%
U\simeq 15.0\Delta $ we are already in the Kondo regime for finite $U$, as
shown in Fig. \ref{Fig33}, where the Kondo peak is well defined. The Fig. %
\ref{Fig13} for $U=20.0\Delta $ corresponds to the symmetrical limit of the
Anderson model, in which the total occupation number is exactly $n_{f}=1.0$.
It is interesting to observe that in this case the atomic approach is able
to reproduce the correct symmetry of the density of states. Finally in Fig. %
\ref{Fig333}, for $U=25.0\Delta $ the Kondo peak continues pinned to the
chemical potential. 
\begin{figure}[th]
\includegraphics[clip,width=0.40%
\textwidth,angle=-90.0]{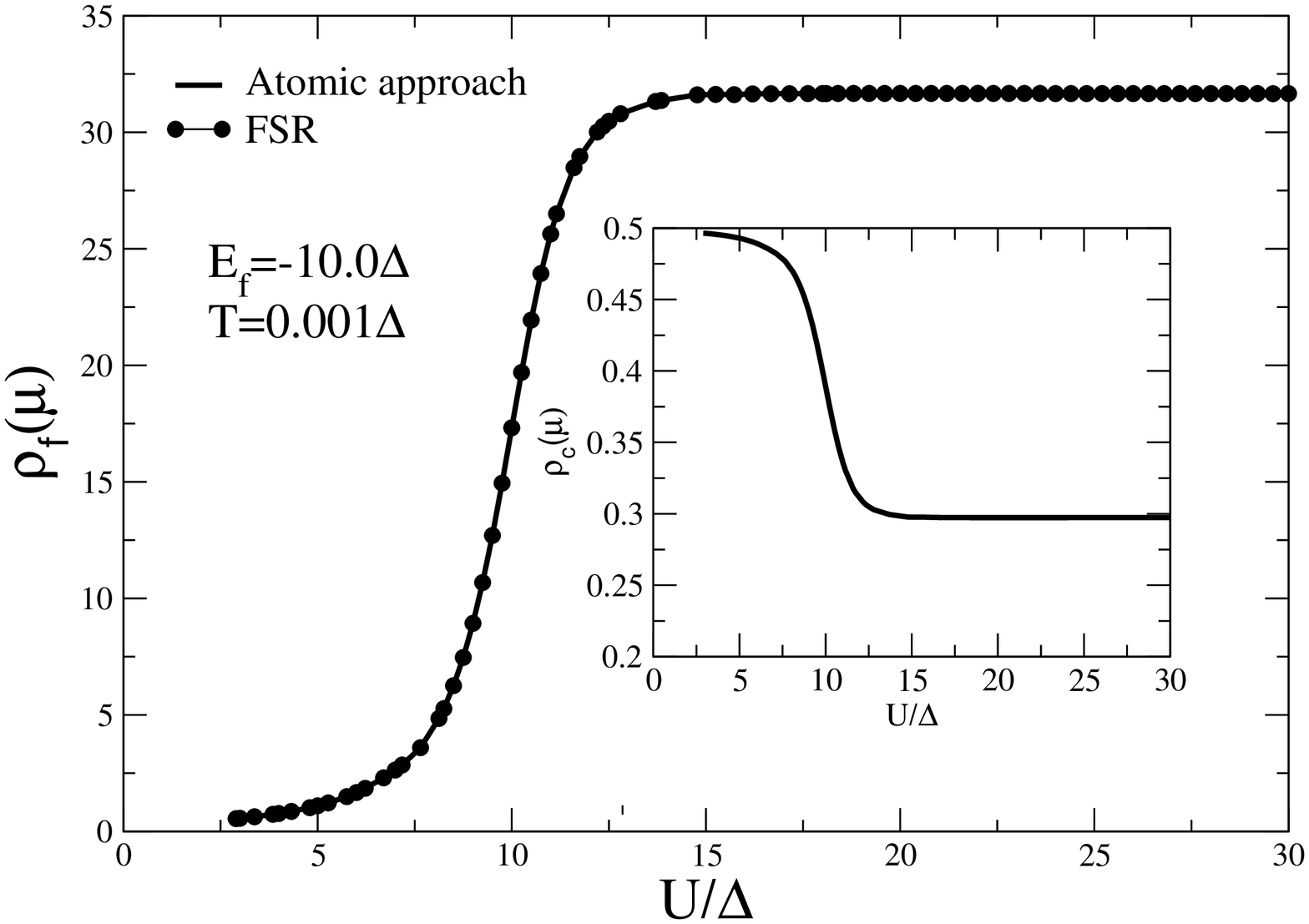}
\caption{Density of states $\protect\rho _{f}(\protect\mu )$ at the chemical
potential $\protect\mu $ as function of the correlation $U$ for $%
E_{f}=-10.0\Delta $ and $T=0.001\Delta $, for the atomic method and for the
Friedel sum rule (FSR). The FSR is represented by points over the curve. In
the inset we represent the density $\protect\rho _{c}(\protect\mu )$ of
conduction electron states at the chemical potential $\protect\mu $, as a
function of the correlation $U$. }
\label{Fig7}
\end{figure}
\begin{figure}[th]
\includegraphics[clip,width=0.40\textwidth,angle=-90.0]{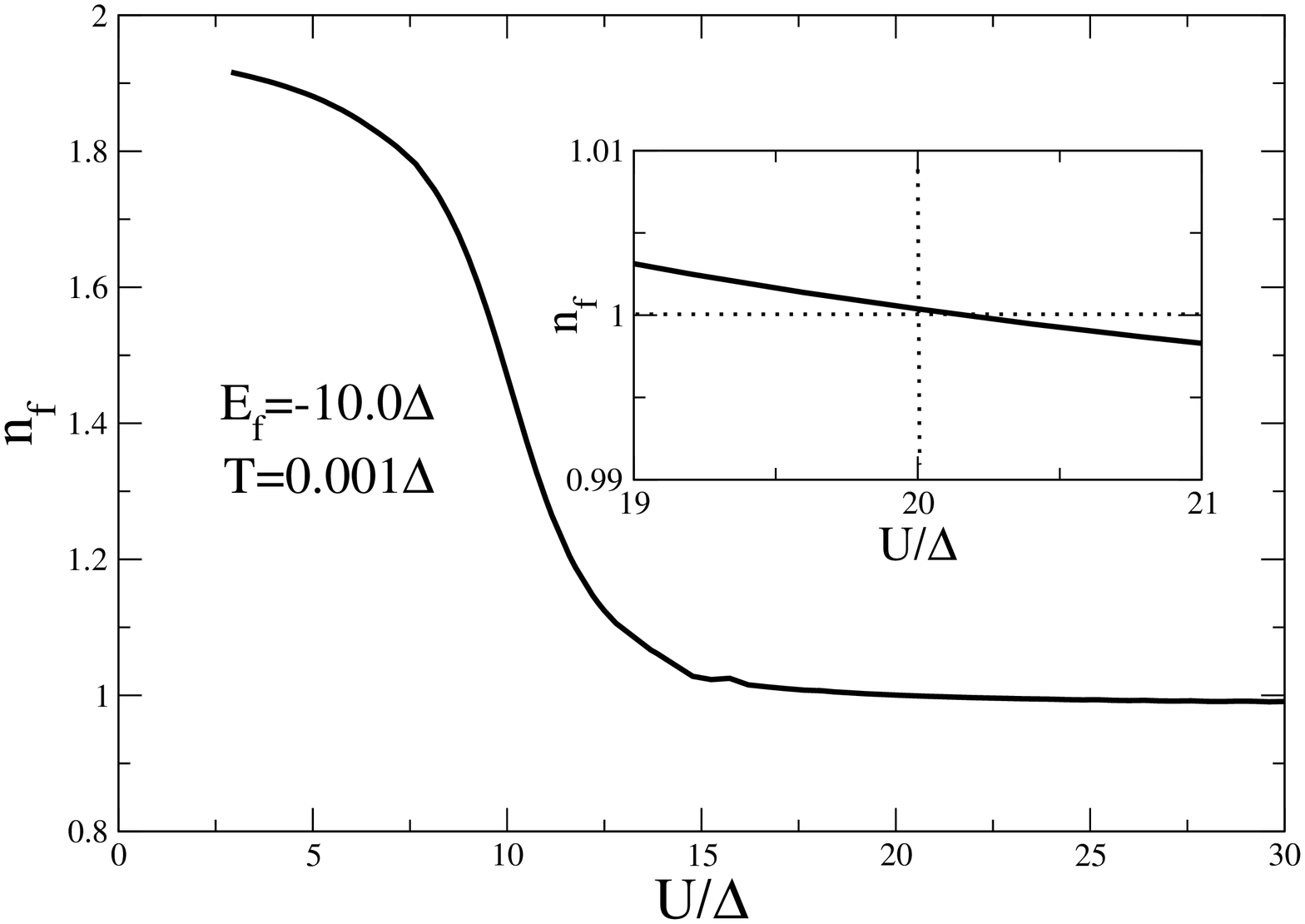}
\caption{The total localized occupation number $n_{f}$ as a function of the
correlation $U$ in $\Delta $ units. In the inset we represent the limit
where the $f$ occupation number becomes exactly $n_{f}=1.0$. This point
corresponds to the symmetrical limit of the Anderson hamiltonian.}
\label{Fig88}
\end{figure}

In Fig. \ref{Fig7} we present the $f$ density of the states at the chemical
potential $\mu $. At the uncorrelated side ($U<15.0\Delta $), the density of
states at the chemical potential is small, but increases as the value of the
correlation $U$ increases. For ($U>15.0\Delta $) we attain the Kondo regime,
where the occupation number $n_{f}\simeq 1$ and the Friedel sum rule
produces ($\rho _{f}(\mu )=1/\pi \Delta $) \cite{Costi86}. In the inset of
the figure we present the density of states of conduction electrons $c$ at
the chemical potential $\mu $, showing a loss of conduction electron states
that migrate to the localized band, screening the impurity and generating
the Kondo effect.

In Fig. \ref{Fig88} we present the total localized occupation number $%
n_{f}=\sum_{\sigma}n_{f,\sigma}$ as a function of the correlation $U$ in $%
\Delta$ units. At the non correlated side ($U < 15.0\Delta$), the $f$
occupation number assume values between $n_{f}=1.0-2.0$. As the correlation $%
U$ increases, at around $U \simeq 15.0\Delta$, the system attains the Kondo
regime and the $n_{f}$ assume values close to the $1.0$ as is expected to
the Kondo limit of the model. In the inset we represent the limit where the $%
f$ occupation number becomes exactly $n_{f}=1.0$. This point corresponds to
the symmetrical limit of the Anderson Hamiltonian.

\section{The different regimes of the model}

\label{sec7}

\begin{figure}[th]
\includegraphics[clip,width=0.40%
\textwidth,angle=-90.0]{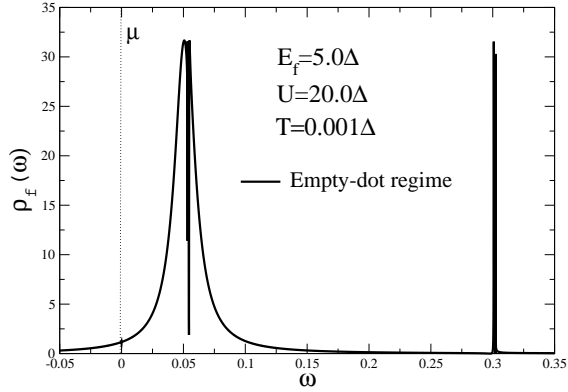}
\caption{Density of states for $T=0.001 \Delta$ , $U=20\Delta$ and $%
E_{f}=5.0 \Delta$. This parameter set corresponds to the empty dot regime.}
\label{Fig107}
\end{figure}

\begin{figure}[th]
\includegraphics[clip,width=0.40%
\textwidth,angle=-90.0]{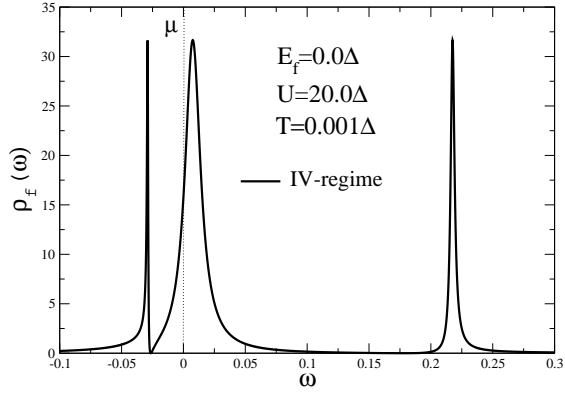}
\caption{Density of states for $T=0.001 \Delta$ , $U=20\Delta$ and $%
E_{f}=0.0 \Delta$. This parameter set corresponds to the intermediate
valence regime (IV).}
\label{Fig11}
\end{figure}

\begin{figure}[th]
\includegraphics[clip,width=0.40%
\textwidth,angle=-90.0]{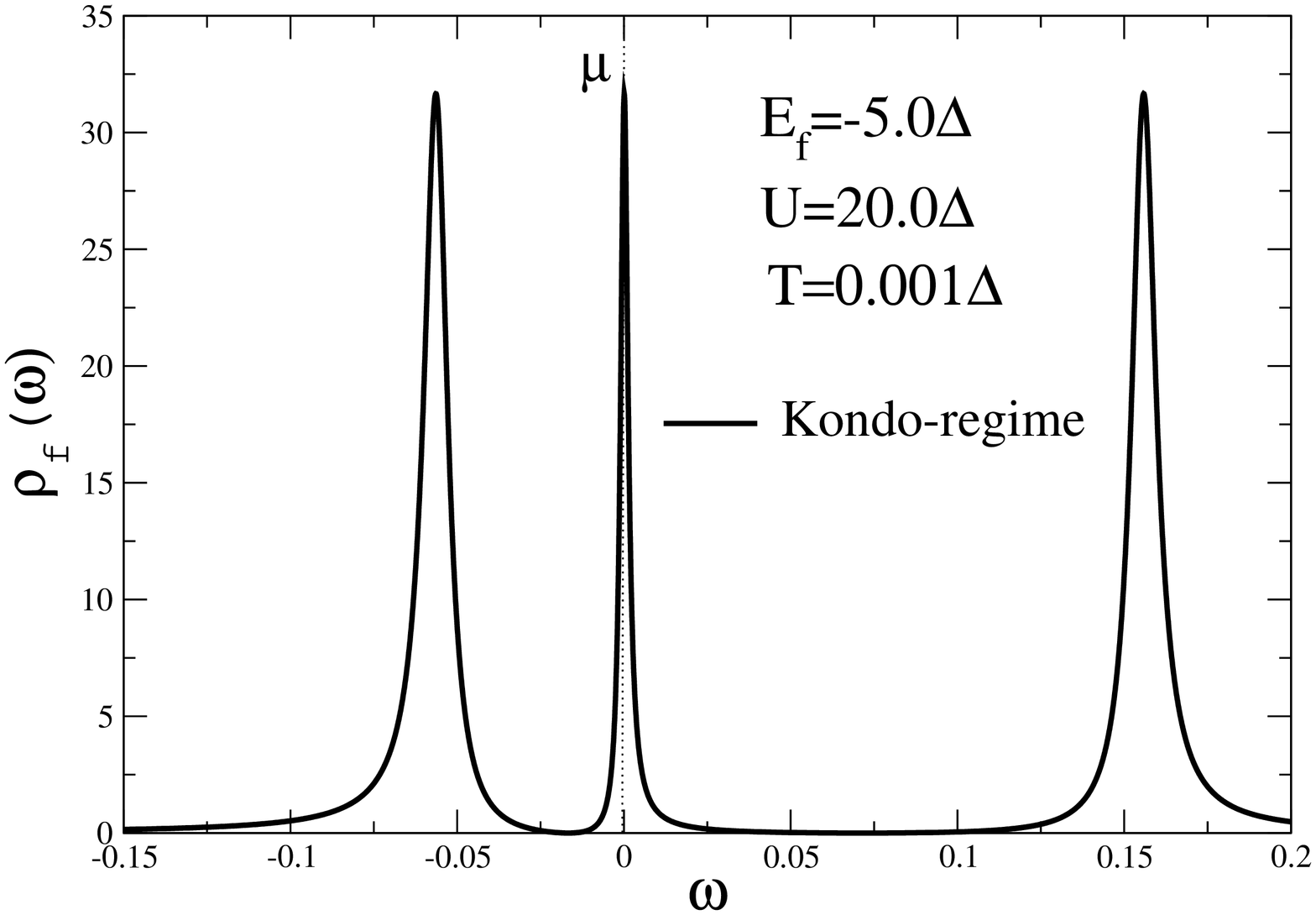}
\caption{Density of states for $T=0.001 \Delta$ , $U=20\Delta$ and $%
E_{f}=-5.0 \Delta$. This parameter set corresponds to the beginning of the
Kondo regime.}
\label{Fig12}
\end{figure}

\begin{figure}[th]
\includegraphics[clip,width=0.40%
\textwidth,angle=-90.0]{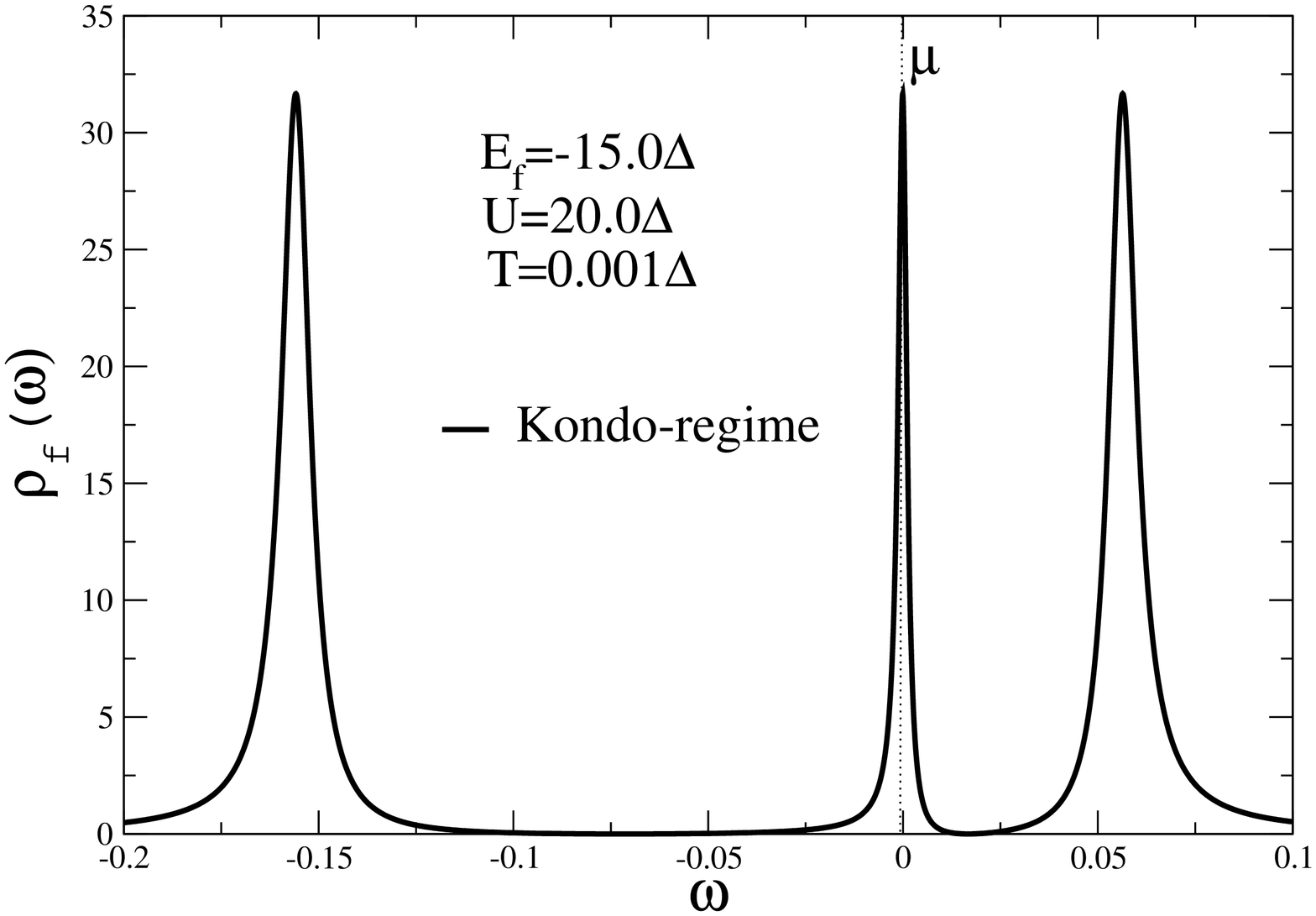}
\caption{Density of states for $T=0.001 \Delta$ , $U=20\Delta$ and $%
E_{f}=-15.0 \Delta$. This parameter set corresponds to the ending of the
Kondo regime.}
\label{Fig14}
\end{figure}

\begin{figure}[th]
\includegraphics[clip,width=0.40%
\textwidth,angle=-90.0]{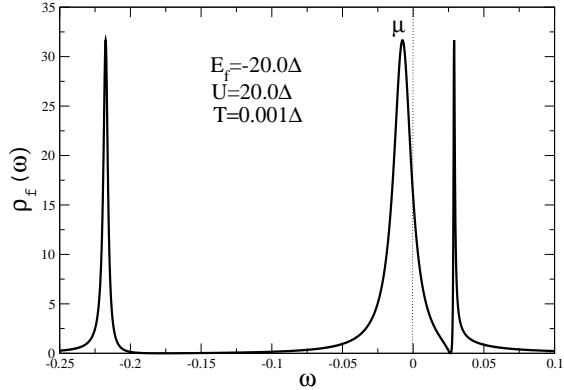}
\caption{Density of states for $T=0.001 \Delta$ , $U=20\Delta$ and $%
E_{f}=-20\Delta$. This region is dominated by the double occupation band and
the $f$ occupation number is greater than one and corresponds to the
crossover from the Kondo to the magnetic region.}
\label{Fig15}
\end{figure}

\begin{figure}[th]
\includegraphics[clip,width=0.40%
\textwidth,angle=-90.0]{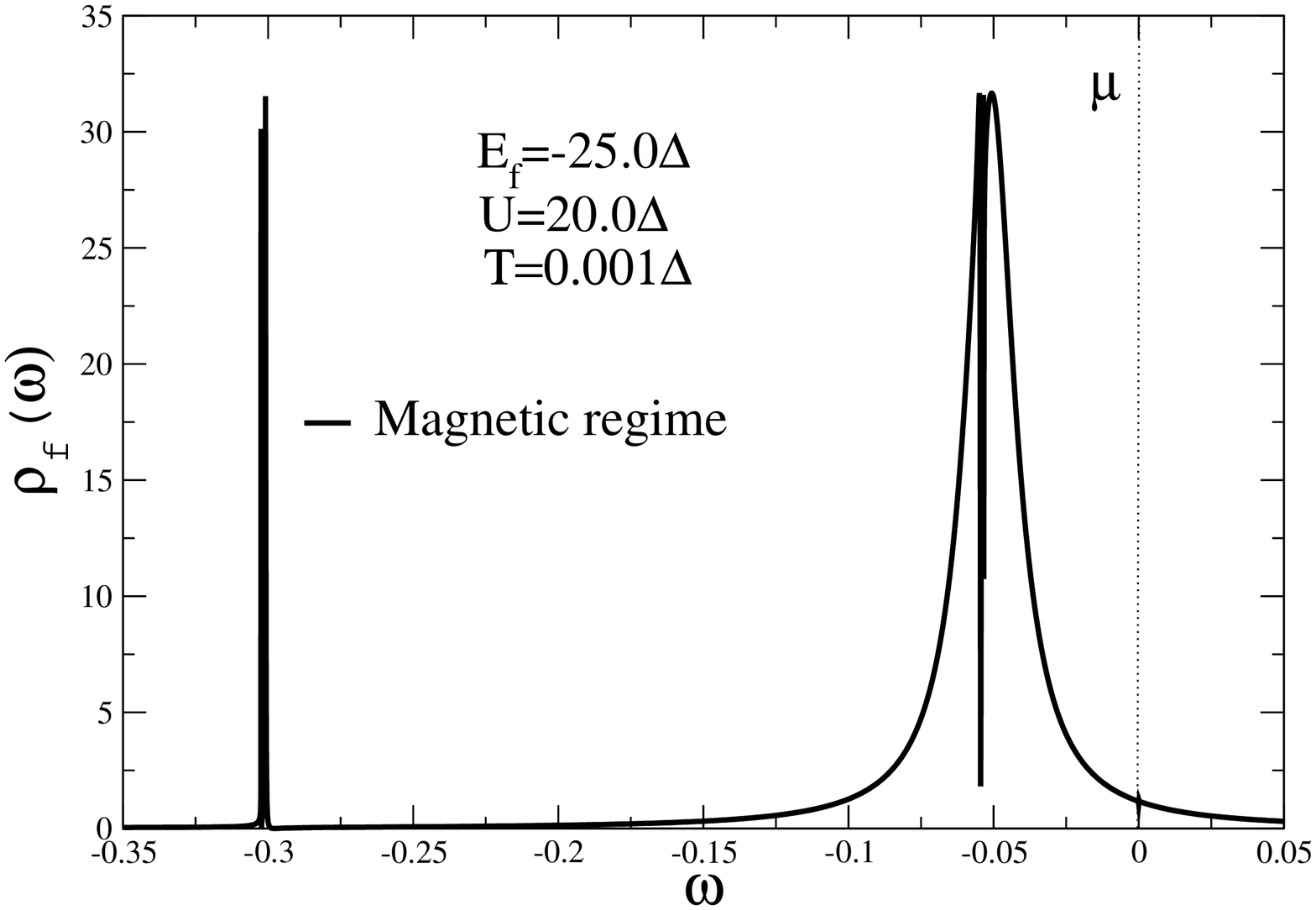}
\caption{Density of states for $T=0.001 \Delta$ , $U=20\Delta$ and $%
E_{f}=-25.0 \Delta$. This limit is completelly dominated by the double
occupation band and the $f$ occupation number is close to $2.0$ and
corresponds the region where $E_{f}-U < 0$.}
\label{Fig16}
\end{figure}

In the set of Figs. (\ref{Fig107}-\ref{Fig16}) we fix the correlation energy
value in $U=20.0\Delta$ and we vary the localized level $E_{f}$ in order to
describe the different regimes of the model: $E_{f}=5.0\Delta$ (empty-dot
regime); $E_{f}=0.0\Delta$ (intermediate valence (IV) regime); $%
E_{f}=-5.0\Delta$ and $E_{f}=-15.0\Delta$ (Kondo regime), $E_{f}=-20.0\Delta$
(crossover from the Kondo to the magnetic regime) and $E_{f}=-25.0\Delta$
(magnetic regime).

In Fig. \ref{Fig107} we represent the empty dot regime. We have only a tail
of the density of states below the chemical potential $\mu$ and the total
occupation number is very low and is given by $n_{f} \simeq 0.12$.

In Fig. \ref{Fig11} we represent a typical intermediate valence situation.
In this case the density of states already exhibit the three peak structures
characteristic of the Kondo regime, but the Kondo peak is not yet formed;
there is a large structure at the chemical potential $\mu$, that generates a
strong charge fluctuation. The total occupation number is characteristic of
the intermediate valence regime; $n_{f} \simeq 0.50$.

In Fig. \ref{Fig12} we represent the beginning of the Kondo regime for $%
E_{f}\simeq -5.0\Delta $ and $n_{f}\simeq 0.96$; the system remain in this
regime until $E_{f}\simeq -0.15\Delta $ and $n_{f}\simeq 1.04$ as indicated
in Fig. \ref{Fig14}. In this region, the Kondo peak is well defined and is
pinned at the chemical potential $\mu $, and in Fig. \ref{Fig17} we present
a resume of all regimes. The interesting point that should be stressed here,
is that the Kondo limit, where the localized occupation number is exactly
equal the unity ($n_{f}=1.0$), is attained in the symmetrical limit of the
Anderson model as indicated in Fig. \ref{Fig13}.

\begin{figure}[th]
\includegraphics[clip,width=0.40%
\textwidth,angle=-90.0]{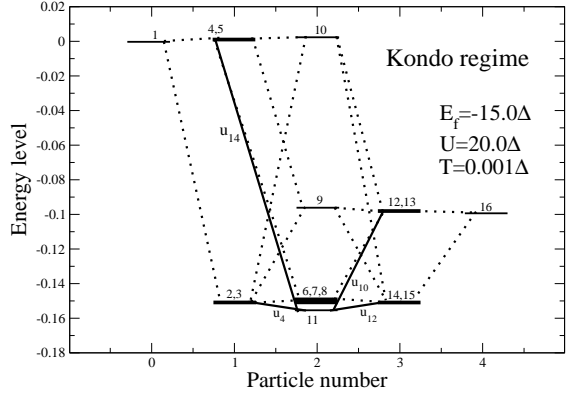}
\caption{The sixteen energy levels $E_{n,r}$ (in units of $\Delta $) of the
AAS,  obtained in Appendix \protect\ref{ApA} (cf. Table III), as a function
of the particle number for typical parameters of the Kondo regime: $%
U=20\Delta $, $E_{f}=-15.0\Delta $ and $T=0.001\Delta $.}
\label{Energy_levels}
\end{figure}

\begin{figure}[th]
\includegraphics[clip,width=0.40\textwidth,angle=-90.0]{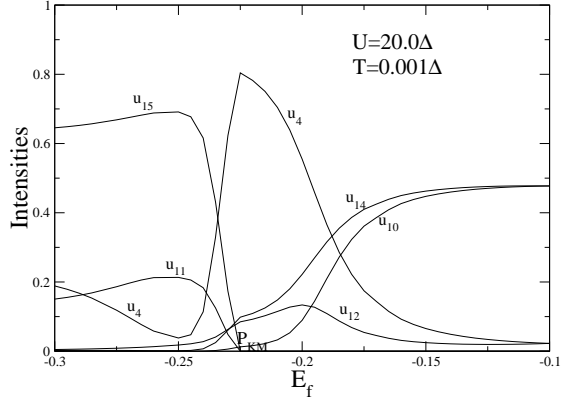}
\caption{Intensities of the residues associating to the transition
\textquotedblleft Kondo-magnetic" as a function of $E_{f}$ (in units of $%
\Delta $), corresponding to $U=20\Delta $, and $T=0.001\Delta $. }
\label{Intensities}
\end{figure}

\begin{figure}[th]
\includegraphics[clip,width=0.40%
\textwidth,angle=-90.0]{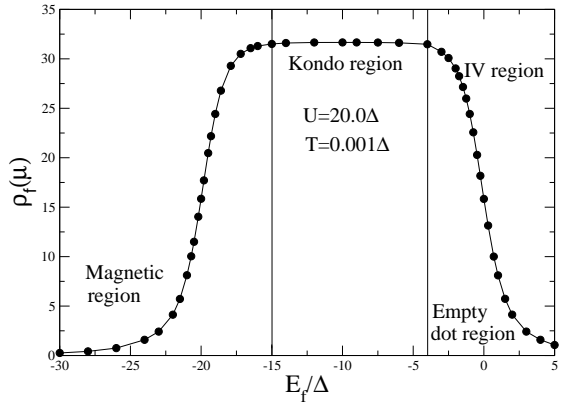}
\caption{Density of states at the chemical potential $\protect\mu$ vs. $%
E_{f} $ for $T=0.001 \Delta$ , $U=20.0\Delta$.}
\label{Fig17}
\end{figure}

For $E_{f}<-15.0\Delta $ the influence of the double occupation band over
the Kondo effect increases more and the Kondo peak enlarges its width as
indicated in the Fig. \ref{Fig15}. In this figure $E_{f}\approx U$ and $%
n_{f}\simeq 1.50$ and there is no Kondo effect. We call this region the
\textquotedblleft magnetic region" because as $E_{f}$ becomes more and more
negative there is a competition between the Kondo state $|11>$ and the two
magnetic states $|14>$,$|15>$ of the atomic solution of the SIAM. We will
discuss this point in more details in Figs. \ref{Energy_levels}-\ref{Fig17}.

Finally in Fig. \ref{Fig16} we represent the limit where $(U + E_{f}) << 0$; in this particular case
the double occupation state is almost completely full and $n_{f} \simeq 1.88$.


It is now convenient to study in more detail the crossover region from the
Kondo to the magnetic regime. In Fig. \ref{Energy_levels} we represent the
sixteen energy levels $E_{n,r}\equiv E_j$ ($j=1, ... ,16)$ of the AAS (cf.
Table III) as a function of the particle number, for the parameters
corresponding to the ending of the Kondo regime as indicated in the Fig. \ref%
{Fig14}: $U=20\Delta $, $E_{f}=-15.0\Delta $ and $T=0.001\Delta $. The lines
joining them are associated to the corresponding poles $u_{i}$ (cf. Table
IV) of the atomic GF in Eq.~(\ref{GAnderson})\, that are given in Eqs.~(\ref%
{E55.97}-\ref{E55.99})

When $E_{f}<0.15$ the double occupation states become available to the
system and there is a competition between the singlet that originates the
Kondo effect represented by the state $|11>$ in Fig. \ref{Energy_levels} and
a magnetic state represented in the same figure by the two degenerate states 
$|14>$,$|15>$.

This point deserves a more detailed study. To do so we consider Fig. \ref%
{Intensities}: at $E_{f}\simeq -22.0\Delta $ there is a change of ground
state from the two-particle Kondo singlet $|11>$ to a three particle
magnetic doublet $(|14>,|15>)$ and we call this point $P_{KM}$ (\textquotedblleft Kondo-magnetic" transition). In that
figure we present the transitions $u_{4},u_{12},u_{10},$ and $u_{14}$ 
connected to the singlet $|11>$, and their intensities are important in the
Kondo region but they decrease strongly when $E_{f}$ becomes closer to the
point $P_{KM}$. For values of $E_{f}<P_{KM}$ the transitions 
$u_{4}$,$u_{11}$ and $u_{15}$ associated to the magnetic state
grow up. This region becomes relevant when the $RKKY$ interaction is
present, like in the problem of several interacting impurities or in the
lattice case. This interaction is associated to the magnetic transitions in
the heavy fermion Kondo problem, in which case this transition is generally
antiferromagnetic.

One interesting problem in which the above discussion applies is the double
quantum dots problem (DQD), which has been recently proposed as a possible
realization of the spin quantum computer \cite{Koppens06}. It constitutes
the minimal system for studying a lattice of magnetic impurities in a
tunable environment. In this system the competition between the Kondo effect
and the RKKY interaction lead to a second-order quantum phase transition
(QPT) but if the electrons can tunnel between impurities this QPT is
replaced by a crossover \cite{Jones88}. 
\begin{figure}[th]
\includegraphics[clip,width=0.40%
\textwidth,angle=-90.0]{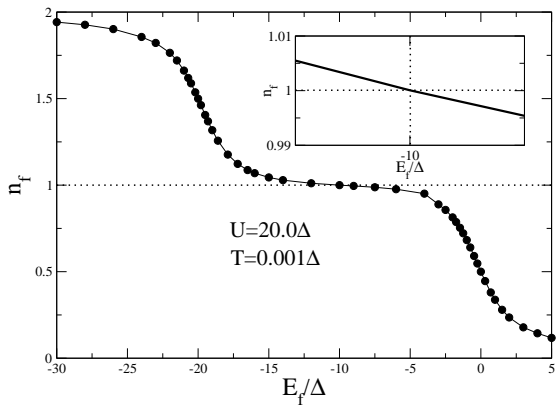}
\caption{The total localized occupation number $n_{f}$ as a function of the
correlation $E_{f}$. In the inset we represent the limit where the $f$
occupation number becomes exactly $n_{f}=1.0$. This point corresponds to the
symmetrical limit of the Anderson Hamiltonian.}
\label{Fig8}
\end{figure}

In Fig. \ref{Fig17} we show the density of states at the chemical potential $%
\mu$ vs. $E_{f}$ for $T=0.001 \Delta$, $U=20\Delta$. The two vertical lines
separate three main regions: intermediate valence to Kondo region and the
Kondo region to the magnetic region. From this result, it is clear that in
the case of finite $U$ the Kondo effect only exists in a limited parameter
region of the SIAM. Fig. \ref{Fig8} shows that the Kondo behavior manifests
itself when the total $f$ occupation number $n_{f}=\sum_{\sigma}n_{f,\sigma}$
assume values close to one $n_{f}=1.0 $; according to the inset of the
figure, this value of $n_{f}$ corresponds to the symmetrical limit of the
SIAM as represented in Fig. \ref{Fig13}.

\section{Conductance of a side-coupled quantum dot for the finite $U$ case}

\label{sec8}

In this Section we apply the atomic method for finite energy correlation $U$%
, to study the electronic transport through a quantum wire (QW) with a
side-coupled quantum dot (QD) \cite{QDnosso}. The quantum dot can be
occupied from zero to two electrons as a function of the chemical potential $%
\mu$. This system has been already studied for $U \rightarrow \infty$ when
the double occupation is forbidden, \cite{Kang2001,QDnosso}. The finite $U$
case is much more realistic, and produces interesting results \cite%
{Seridonio09} for the conductance that should be compared with recent
experimental data \cite{Kobayashy2004,Masahiro05,Fuhrer06}.

\begin{figure}[tbh]
\includegraphics[clip,width=0.45\textwidth,angle=0.]{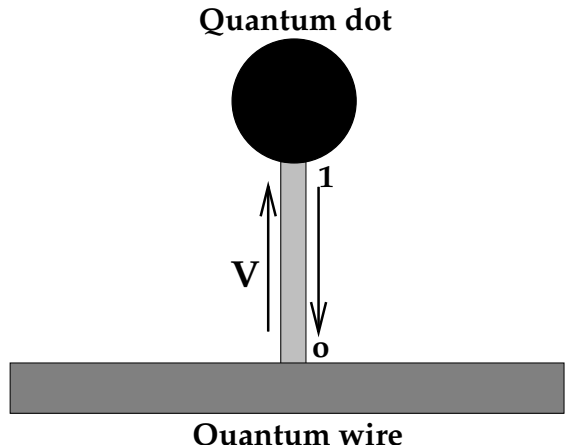}
\caption{Pictorial view of a quantum dot side-coupled to a conduction
channel. }
\label{Lateral_dot}
\end{figure}

In Fig. \ref{Lateral_dot} we present a pictorial view of the quantum dot,
side-coupled to a ballistic channel.

Electron transport is coherent at low temperature and bias voltage, and a
linear-response conductance is given by the Landauer-type formula \cite%
{Kang2001} 
\begin{equation}
G=\frac{2e^{2}}{\hbar}\int\left( -\frac{\partial n_{f}}{\partial\omega }%
\right) S(\omega)d\omega,
\end{equation}
where $n_{F}$ is the Fermi function and $S(\omega)$ is the transmission
probability of an electron with energy $\hbar\omega$. This probability is
given by 
\begin{equation}
S(\omega)=\Gamma^{2}\mid G_{00}^{\sigma}\mid^{2},
\end{equation}
where $\Gamma$ corresponds to the coupling strength of the site $0$ to the
wire. $G_{00}^{\sigma}$ can be calculated by the Dyson equation, where $%
\tilde{V}=|0\left\rangle V\right\langle 1|+|1\left\rangle V\right\langle 0|$
is the hybridization. The dressed Green's functions at the site 0 can be
written in terms of the undressed localized Green's functions $g_{11}$ at
the QD and of the undressed Green's functions $g_{00}$ of the conduction
electrons: 
\begin{equation}
G_{00}^{\sigma}=g_{00}^{\sigma}+g_{00}^{\sigma}VG_{10}^{\sigma}+g_{01}^{%
\sigma}VG_{00}^{\sigma},
\end{equation}%
\begin{equation}
G_{10}^{\sigma}=g_{10}^{\sigma}+g_{10}^{\sigma}VG_{10}^{\sigma}+g_{11}^{%
\sigma}VG_{00}^{\sigma}.
\end{equation}
Solving this system of equations, and taking into account that the
non-diagonal bare conduction Green's functions vanish: $g^{\sigma}_{10}=0$
and $g^{\sigma}_{01}=0$, we can write 
\begin{equation}
G_{00}^{\sigma}=\frac{g_{00}^{\sigma}}{(1+g_{00}^{\sigma}V^{2}g_{11}^{\sigma
})},
\end{equation}
where 
\begin{equation}
g_{00}^{\sigma}=\left( \frac{-1}{2D}\right) \ln\left( \frac{z+D+\mu }{z-D+\mu%
}\right) \hspace{0.4cm};\hspace{0.4cm}g_{11}^{\sigma}=M_{\uparrow }^{at}(z),
\label{G0c}
\end{equation}
and $M_{\uparrow}^{at}(z)$ is obtained employing the atomic approach Green's
functions given by Eq. (\ref{5.121})

\begin{figure}[th]
\includegraphics[clip,width=0.40\textwidth,angle=-90.]{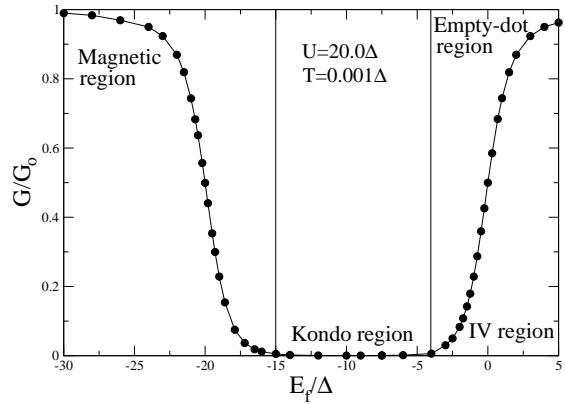}
\caption{Conductance of the side-coupled quantum dot for $U=20.0\Delta$ and $%
T=0.001\Delta$.}
\label{condutancia}
\end{figure}

In Fig. \ref{condutancia} we calculate the conductance corresponding to the
parameters employed in Sec. \ref{sec6}. We show the conductance as a
function of $E_{f}$ for $U=20\Delta$ and at temperature $T=0.001\Delta$. We
can observe four important regions indicated in the graph. The first is the
empty-dot region. In this situation, the value of the localized level $E_{f}$
is positive and is located above the chemical potential $\mu$; the localized
occupation number goes to zero $(n_{f}\longrightarrow 0)$ and the
conductance goes to one $(G/G_{0}\longrightarrow 1)$. In this situation
there is practically no destructive interference between the dot and the
wire.

The next regime is the intermediate valence, which is indicated in the graph
by ``IV region", and the localized level $E_{f}$ is located around the
chemical potential $\mu=0$. In this region the value of the $f$ occupation
number varies strongly as a consequence of the charge fluctuation, the
quantum interference effects start to increase and the Kondo peak appears
when $n_{f} \rightarrow 1$.

The next region in the graph shows the effect of the Kondo regime in the
side-coupled QD system. The occupation number is around one $(n_{f} \simeq
1) $ and the conductance goes to zero $(G/G_{0}\longrightarrow 0)$. The
Kondo peak is formed, and there is a perfect destructive quantum
interference of $\pi/2$ between the electrons that go around the wire and
those that ``visit" the QD and return to the wire.

In the last region the $f$ occupation is greater than one, and the double
occupation dominates this regime as indicated in Figs. \ref{Fig15}-\ref%
{Fig16}. This is the ``magnetic" region and we can observe a competition
between the Kondo effect and the magnetic state, as discussed in the text of
Figs. \ref{Energy_levels}-\ref{Fig17}. The occupation number goes to two $%
(n_{f}\longrightarrow 2)$ as the $E_{f}$ becomes more negative and the
conductance goes again to one $(G/G_{0} \longrightarrow 1)$.

\section{Conclusions}

\label{sec9}

Employing the atomic approach, we characterized well the formation of the
Kondo peak for all the parameters of the SIAM by calculating several curves
of density of states, varying the value of the coulomb repulsion $U$, as
well as the position of the localized impurity level $E_{f}$.

As a general conclusion we can say that we have developed a general and
simple method to calculate the low temperature properties of the Anderson
impurity model, and we call this method the atomic approach, because the
starting point of the method is the zero conduction bandwidth limit of the
Anderson impurity model. The approach is extremely simple, satisfies the
Friedel sum rule by construction and is valid, at low temperatures, for all
the relevant range of parameters and for all the coupling regimes of the
SIAM, namely for the weak, intermediate and strong correlation regimes of
the model.

The atomic approach is a good candidate to describe nanoscopic systems with
correlated electrons that present Kondo effect, and it produces excellent
results for the occupation numbers and for the dynamical properties. Due to
its simplicity, low computational cost (the computational time consuming to
obtain a density of states curve takes few seconds or less) the atomic
approach could be applied to study nanoscopic correlated systems, like
quantum dots.

\begin{acknowledgments}
We would like to express our gratitude to the Conselho Nacional de
Desenvolvimento Cientifico (CNPq) - Brazil.
\end{acknowledgments}

\appendix

\section{The atomic Green functions}

\label{ApA}

\begin{table}[tbh]
\begin{center}
\begin{equation*}
\begin{array}{|l|c|c|c|}
\hline
|m,\sigma> \hspace{0.5cm} & E \hspace{0.5cm} & n \hspace{0.5cm} & S_{z} 
\hspace{0.5cm} \\ \hline
|0,0> \hspace{0.5cm} & 0 \hspace{0.5cm} & 0 \hspace{0.5cm} & 0 \hspace{0.5cm}
\\ \hline
|+,0> \hspace{0.5cm} & \varepsilon_{f} \hspace{0.5cm} & 1 \hspace{0.5cm} & 
1/2 \hspace{0.5cm} \\ \hline
|0,\uparrow> \hspace{0.5cm} & \varepsilon_{q} \hspace{0.5cm} & 1 \hspace{%
0.5cm} & 1/2 \hspace{0.5cm} \\ \hline
|-,0> \hspace{0.5cm} & \varepsilon_{f} \hspace{0.5cm} & 1 \hspace{0.5cm} & 
-1/2 \hspace{0.5cm} \\ \hline
|0,\downarrow> \hspace{0.5cm} & \varepsilon_{q} \hspace{0.5cm} & 1 \hspace{%
0.5cm} & -1/2 \hspace{0.5cm} \\ \hline
|+,\uparrow > \hspace{0.5cm} & \varepsilon_{f}+\varepsilon_{q} \hspace{0.5cm}
& 2 \hspace{0.5cm} & 1 \hspace{0.5cm} \\ \hline
|-,\downarrow> \hspace{0.5cm} & \varepsilon_{f}+\varepsilon_{q} \hspace{0.5cm%
} & 2 \hspace{0.5cm} & -1 \hspace{0.5cm} \\ \hline
|+,\downarrow > \hspace{0.5cm} & \varepsilon_{f}+\varepsilon_{q} \hspace{%
0.5cm} & 2 \hspace{1cm} & 0 \hspace{0.5cm} \\ \hline
|-,\uparrow>\hspace{0.5cm} & \varepsilon_{f}+\varepsilon_{q} \hspace{0.5cm}
& 2 \hspace{1cm} & 0 \hspace{1cm} \\ \hline
|0,\uparrow\downarrow> \hspace{0.5cm} & 2\varepsilon_{q} \hspace{0.5cm} & 2 
\hspace{0.5cm} & 0 \hspace{0.5cm} \\ \hline
|d,0> \hspace{0.5cm} & 2\varepsilon_{f}+U \hspace{0.5cm} & 2 \hspace{1cm} & 
0 \hspace{1cm} \\ \hline
|d,\uparrow> \hspace{0.5cm} & 2\varepsilon_{f}+\varepsilon_{q}+U \hspace{%
0.5cm} & 3 \hspace{0.5cm} & 1/2 \hspace{0.5cm} \\ \hline
|+,\uparrow\downarrow> \hspace{0.5cm} & \varepsilon_{f}+2\varepsilon_{q} 
\hspace{0.5cm} & 3 \hspace{0.5cm} & 1/2 \hspace{0.5cm} \\ \hline
|d,\downarrow> \hspace{0.5cm} & 2\varepsilon_{f}+\varepsilon_{q}+U \hspace{%
0.5cm} & 3 \hspace{0.5cm} & -1/2 \hspace{0.5cm} \\ \hline
|-,\uparrow\downarrow> \hspace{0.5cm} & \varepsilon_{f}+2\varepsilon_{q} 
\hspace{0.5cm} & 3 \hspace{0.5cm} & -1/2 \hspace{0.5cm} \\ \hline
|d,\uparrow\downarrow> \hspace{0.5cm} & 2\varepsilon_{f}+2\varepsilon_{q}+U 
\hspace{0.5cm} & 4 \hspace{0.5cm} & 0 \hspace{0.5cm} \\ \hline
\end{array}%
\end{equation*}%
\end{center}
\caption[TABLE I]{States of the Anderson impurity in the limit $2D=V=0$. The
columns indicates the states $|n,\protect\sigma >$, the energies $E_{o}$,
the number of electrons $n$ and the spin component $S_{z}$.\newline
}
\label{Table1}
\end{table}
In this Appendix we present the details of the calculation of the exact
Green's function of the Anderson impurity model in the limit of zero
conduction bandwidth. In this limit all the hoping contributions are
eliminated from the Hamiltonian, because we also take $V_{f,\mathbf{k}%
,\sigma }=V$ in Eqs. (\ref{E5.1}-\ref{E5.2}), i.e.: a local hybridization.
Transforming the conduction electrons to the Wannier representation we then
have an independent system at each site of the crystal. In this limit there
is an isolated metal atom at each site, one of them being the Anderson
impurity, and the system Hamiltonian can be diagonalized exactly. In the
Anderson Hamiltonian, there are four possible occupations $(0,\uparrow
,\downarrow ,\uparrow \downarrow )$ of the conduction electrons at any site,
and at the impurity there are also four possible occupations of the local $f$
electrons: $(0,+\frac{1}{2},-\frac{1}{2},d)$. At the impurity site we then
have a Fock space with sixteen states characterized by $|m,\sigma >$ as
shown in Table \ref{Table1}, and the hybridization mixes states with equal
particle number and with the same spin component $z$. We should then
diagonalize a $16$x$16$ matrix which presents a block structure that
simplifies the calculation, because the greater block is $3$x$3$ and we
employ Cardano's formula to solve the associated third degree algebraic
equation. The results of this calculation are presented in Table \ref{Table2}%
. 
\begin{widetext}
\begin{table*}[tbh]
\begin{center}
\begin{equation*}
\begin{array}{|l|l|c|c|}
\hline
Eigenstates\hspace{0.3cm}|j>\equiv |n,r> & Eigenvalues\text{ }E_{j}\equiv
E_{n,r}\hspace{0.3cm} & n\hspace{0.3cm} & S_{z} \\ \hline
|1>\equiv |0,1>=|0,0\rangle \hspace{0.3cm} & E_{1}\equiv E_{0,1}=0\hspace{%
0.3cm} & 0\hspace{0.3cm} & 0 \\ \hline
|2\rangle \equiv |1,1>=\cos \phi |+,0\rangle -\sin \phi |0,\uparrow \rangle
\hspace{0.3cm} & E_{2}\equiv E_{1,1}=\frac{1}{2}(\varepsilon
_{f}+\varepsilon _{q}-\Delta )\hspace{0.3cm} & 1\hspace{0.3cm} & +\frac{1}{2}
\\ \hline
|3\rangle \equiv |1,2>=\cos \phi |-,0\rangle -\sin \phi |0,\downarrow
\rangle \hspace{0.3cm} & E_{3}\equiv E_{1,2}=E_{2}\hspace{0.3cm} & 1\hspace{%
0.3cm} & -\frac{1}{2} \\ \hline
|4\rangle \equiv |1,3>=\sin \phi |+,0\rangle +\cos \phi |0,\uparrow \rangle
\hspace{0.3cm} & E_{4}\equiv E_{1,3}=\frac{1}{2}(\varepsilon
_{f}+\varepsilon _{q}+\Delta )\hspace{0.3cm} & 1\hspace{0.3cm} & +\frac{1}{2}
\\ \hline
|5\rangle \equiv |1,4>=\sin \phi |-,0\rangle +\cos \phi |0,\downarrow
\rangle \hspace{0.3cm} & E_{5}\equiv E_{1,4}=E_{4}\hspace{0.3cm} & 1\hspace{%
0.3cm} & -\frac{1}{2} \\ \hline
|6\rangle \equiv |2,1>=|+,\uparrow \rangle \hspace{0.3cm} & E_{6}\equiv
E_{2,1}=\varepsilon _{f}+\varepsilon _{q}\hspace{0.3cm} & 2\hspace{0.3cm} & 1
\\ \hline
|7\rangle \equiv |2,2>=|-,\downarrow \rangle \hspace{0.3cm} & E_{7}\equiv
E_{2,2}=E_{6}\hspace{0.3cm} & 2\hspace{0.3cm} & -1 \\ \hline
|8\rangle \equiv |2,3>=1/\sqrt{2}(|+,\downarrow \rangle +|-,\uparrow \rangle
)\hspace{0.3cm} & E_{8}\equiv E_{2,3}=E_{7}\hspace{0.3cm} & 2\hspace{0.3cm}
& 0 \\ \hline
|9\rangle \equiv |2,4>=a_{9}[|+,\downarrow \rangle -|-,\uparrow \rangle
]+b_{9}|d,0\rangle +c_{9}|0,\uparrow \downarrow \rangle \hspace{0.3cm} &
E_{9}\equiv E_{2,4}=2\sqrt{-Q}\cos (\frac{\theta _{1}}{3})\hspace{0.3cm} & 2%
\hspace{0.3cm} & 0 \\ \hline
|10\rangle \equiv |2,5>=a_{10}[|+,\downarrow \rangle -|-,\uparrow \rangle
]+b_{10}|d,0\rangle +c_{10}|0,\uparrow \downarrow \rangle \hspace{0.3cm} &
E_{10}\equiv E_{2,5}=2\sqrt{-Q}\cos (\frac{\theta _{1}+2\pi }{3})\hspace{%
0.3cm} & 2\hspace{0.3cm} & 0 \\ \hline
|11\rangle \equiv |2,6>=a_{11}[|+,\downarrow \rangle -|-,\uparrow \rangle
]+b_{11}|d,0\rangle +c_{11}|0,\uparrow \downarrow \rangle \hspace{0.3cm} &
E_{11}\equiv E_{2,6}=2\sqrt{-Q}\cos (\frac{\theta _{1}+4\pi }{3})\hspace{%
0.3cm} & 2\hspace{0.3cm} & 0 \\ \hline
|12\rangle \equiv |3,1>=\sin (\theta )|d,\uparrow \rangle +\cos (\theta
)|+,\uparrow \downarrow \rangle \hspace{0.3cm} & E_{12}\equiv E_{3,1}=\frac{1%
}{2}(3\varepsilon _{f}+3\varepsilon _{q}+U+\Delta ^{\prime })\hspace{0.3cm}
& 3\hspace{0.3cm} & +\frac{1}{2} \\ \hline
|13\rangle \equiv |3,2>=\sin (\theta )|d,\downarrow \rangle +\cos (\theta
)|-,\uparrow \downarrow \rangle \hspace{0.3cm} & E_{13}\equiv E_{3,2}=E_{12}%
\hspace{0.3cm} & 3\hspace{0.3cm} & -\frac{1}{2} \\ \hline
|14\rangle \equiv |3,3>=-\cos (\theta )|d,\uparrow \rangle +\sin (\theta
)|+,\uparrow \downarrow \rangle \hspace{0.3cm} & E_{14}\equiv E_{3,3}=\frac{1%
}{2}(3\varepsilon _{f}+3\varepsilon _{q}+U-\Delta ^{\prime })\hspace{0.3cm}
& 3\hspace{0.3cm} & +\frac{1}{2} \\ \hline
|15\rangle \equiv |3,4>=-\cos (\theta )|d,\downarrow \rangle +\sin (\theta
)|-,\uparrow \downarrow \rangle \hspace{0.3cm} & E_{15}\equiv E_{3,4}=E_{14}%
\hspace{0.3cm} & 3\hspace{0.3cm} & -\frac{1}{2} \\ \hline
|16\rangle \equiv |4,1>=|d,\uparrow \downarrow \rangle \hspace{0.3cm} &
E_{16}\equiv E_{4,1}=2\varepsilon _{f}+2\varepsilon _{q}+U\hspace{0.3cm} & 4%
\hspace{0.3cm} & 0 \\ \hline
\end{array}%
\end{equation*}
\end{center}%
\caption[TABLE III]{ Exact solution (AAS) of the Anderson model in the limit
of local hybridization and conduction band with zero width. The sixteen
eigenstates $\mid n,r\rangle $ have energies $E_{n,r}$, where $n$ is the
number of electrons and $S_{z}$ is the spin component, and we abbreviate $%
|j>\equiv |n,r>$ and $E_{j}\equiv E_{n,r}$ $(j=1,\ldots ,16)$.We use $%
\varepsilon _{n,r}=E_{n,r}-n\mu $, and the other parameters are given by
\newline
\noindent $\Delta=[(\varepsilon_{f}- \varepsilon_{q})^{2}+4V^{2}]^{1/2}$; $%
\Delta^{\prime}=[(\varepsilon_{f}+U- \varepsilon_{q})^{2}+4V^{2}]^{1/2}$;
\newline
$tg\phi=2V/(\varepsilon_{q}- \varepsilon_{f}+\Delta)$; $tg\theta=2V/(%
\varepsilon_{f}+U- \varepsilon_{q}-\Delta^{\prime})$; \newline
$a_{i}=\frac{1}{\sqrt{2+4V^{2}[(E_{i}-2\varepsilon_{f}-U)^{-2}+(E_{i}-2%
\varepsilon_{q})^{-2}]}}$; $b_{i}=\frac{2V}{E_{i}-2\varepsilon_{f}-U}a_{i}$;
$c_{i}=\frac{2V}{E_{9}-2\varepsilon_{q}}a_{i}$; $i=9,10,11$; \newline
$\theta_{1}=\arccos\frac{R}{\sqrt{(-Q)^{3}}}$; $Q=-\frac{1}{9}%
[12V^{2}+(\varepsilon_{q}+
\varepsilon_{f})^{2}+(2\varepsilon_{f}+U)^{2}+(2\varepsilon_{q})^{2}-
(\varepsilon_{q}+\varepsilon_{f})(2\varepsilon_{f}+U)-(\varepsilon_{q}+
\varepsilon_{f})(2\varepsilon_{q})-(2\varepsilon_{f}+U)(2\varepsilon_{q})]$;%
\newline
$R=\frac{1}{54}\{-3[(\varepsilon_{q}+
\varepsilon_{f})^{2}(2\varepsilon_{f}+U)+(\varepsilon_{q}+
\varepsilon_{f})^{2}(2\varepsilon_{q})+(2\varepsilon_{f}+U)^{2}(%
\varepsilon_{q}+
\varepsilon_{f})+(2\varepsilon_{f}+U)^{2}(2\varepsilon_{q})+(2%
\varepsilon_{q})^{2}(\varepsilon_{q}+ \varepsilon_{f})+
(2\varepsilon_{q})^{2}(2\varepsilon_{f}+U)]+12(\varepsilon_{q}+
\varepsilon_{f})(2\varepsilon_{f}+U)(2\varepsilon_{q})+18V^{2}[2(%
\varepsilon_{q}+
\varepsilon_{f})-(2\varepsilon_{f}+U)-(2\varepsilon_{q})]+2[(%
\varepsilon_{q}+
\varepsilon_{f})^{3}+(2\varepsilon_{f}+U)^{3}+(2\varepsilon_{q})^{3}]\}$. }%
\label{Table2}
\end{table*}
\end{widetext}

To obtain the localized atomic Green's functions of the impurity in the zero
width limit, we use Zubarev's \cite{Zubarev60} equation

\begin{equation}
\mathcal{G}_{\alpha\alpha^{\prime}}^{ff,at}(i\omega_{s})=-e^{\beta\Omega}%
\sum_{n,r,r^{\prime}}\frac{\exp(-\beta\varepsilon_{n-1,r})+\exp(-\beta
\varepsilon_{n,r^{\prime}})}{i\omega_{s}+\varepsilon_{n-1,r}-\varepsilon
_{n,r^{\prime}}}\times  \notag
\end{equation}
\begin{equation}
\times \left\langle n-1,r\right\vert\ X_{j,\alpha}\left\vert
n,r^{\prime}\right\rangle \left\langle n,r^{\prime}\right\vert \ X_{j,\alpha
^{\prime}}^{\dagger}\ \left\vert n-1,r\right\rangle ,  \label{Eq3.11}
\end{equation}
where $\Omega$ is the thermodynamical potential and the eigenvalues $E_{nj}$
and eigenvectors $|nj>$ correspond to the complete solution of the
Hamiltonian. The final result is the following%
\begin{equation}  \label{GAnderson}
G^{ff,at}(\omega) = e^{\beta\Omega} \sum_{i=1}^{16} \frac{m_{i}}{\omega-u_{i}%
},
\end{equation}
where the $u_{i}$ poles of the Green's functions are given in Table \ref%
{Table4} \newline

\begin{table}[tbh]
\begin{center}
\begin{tabular}{|ll|}
\hline
$u_{1}=$ & $E_{2}-E_{1}=E_{6}-E_{4}$ \\ \hline
$u_{2}=$ & $E_{5}-E_{1}=E_{6}-E_{2}$ \\ \hline
$u_{3}=$ & $E_{10}-E_{2}$ \\ \hline
$u_{4}=$ & $E_{11}-E_{2}$ \\ \hline
$u_{5}=$ & $E_{9}-E_{2}$ \\ \hline
$u_{6}=$ & $E_{10}-E_{4}$ \\ \hline
$u_{7}=$ & $E_{12}-E_{6}=E_{16}-E_{14}$ \\ \hline
$u_{8}=$ & $E_{12}-E_{9}$ \\ \hline
$u_{9}=$ & $E_{12}-E_{10}$ \\ \hline
$u_{10}=$ & $E_{12}-E_{11}$ \\ \hline
$u_{11}=$ & $E_{14}-E_{10}$ \\ \hline
$u_{12}=$ & $E_{14}-E_{11}$ \\ \hline
$u_{13}=$ & $E_{9}-E_{4}$ \\ \hline
$u_{14}=$ & $E_{11}-E_{4}$ \\ \hline
$u_{15}=$ & $E_{14}-E_{6}=E_{16}-E_{12}$ \\ \hline
$u_{16}=$ & $E_{14}-E_{9}$ \\ \hline
\end{tabular}%
\end{center}
\caption{Poles of the Green's functions, corresponding to all the possible
transitions in the atomic solution.\newline
}
\label{Table4}
\end{table}
The residues for the localized electrons are given by \newline
\noindent $m_{1}=cos(\phi )^{2}[e^{-\beta E_{1}}+e^{-\beta E_{2}}+\frac{3}{2}%
e^{-\beta E_{4}}+\frac{3}{2}e^{-\beta E_{6}}]$, \newline
$m_{2}=sin(\phi )^{2}[e^{-\beta E_{1}}+e^{-\beta E_{4}}+\frac{3}{2}e^{-\beta
E_{2}}+\frac{3}{2}e^{-\beta E_{6}}]$, \newline
$m_{3}=(e^{-\beta E_{3}}+e^{-\beta E_{10}})[(a_{10}sin(\phi
))^{2}+(b_{10}cos(\phi ))^{2}]$, \newline
$m_{4}=(e^{-\beta E_{3}}+e^{-\beta E_{11}})[(a_{11}sin(\phi
))^{2}+(b_{11}cos(\phi ))^{2}]$, \newline
$m_{5}=(e^{-\beta E_{3}}+e^{-\beta E_{9}})[(a_{9}sin(\phi
))^{2}+(b_{9}cos(\phi ))^{2}]$, \newline
$m_{6}=(e^{-\beta E_{4}}+e^{-\beta E_{10}})[(a_{10}cos(\phi
))^{2}+(b_{10}sin(\phi ))^{2}]$, \newline
$m_{7}=sin(\theta )^{2}[\frac{3}{2}(e^{-\beta E_{8}}+e^{-\beta
E_{12}})+(e^{-\beta E_{15}}+e^{-\beta E_{16}})]$, \newline
$m_{8}=(e^{-\beta E_{9}}+e^{-\beta E_{12}})[(c_{9}cos(\theta
))^{2}+(a_{9}sin(\theta ))^{2}]$, \newline
$m_{9}=(e^{-\beta E_{10}}+e^{-\beta E_{12}})[(c_{10}cos(\theta
))^{2}+(a_{10}sin(\theta ))^{2}]$, \newline
$m_{10}=(e^{-\beta E_{11}}+e^{-\beta E_{12}})[(c_{11}cos(\theta
))^{2}+(a_{11}sin(\theta ))^{2}]$, \newline
$m_{11}=(e^{-\beta E_{10}}+e^{-\beta E_{14}})[(c_{10}sin(\theta
))^{2}+(a_{10}cos(\theta ))^{2}]$, \newline
$m_{12}=(e^{-\beta E_{11}}+e^{-\beta E_{15}})[(c_{11}sin(\theta
))^{2}+(a_{11}cos(\theta ))^{2}]$, \newline
$m_{13}=(e^{-\beta E_{5}}+e^{-\beta E_{9}})[(a_{9}cos(\phi
))^{2}+(b_{9}sin(\phi ))^{2}]$, \newline
$m_{14}=(e^{-\beta E_{5}}+e^{-\beta E_{11}})[(a_{11}cos(\phi
))^{2}+(b_{11}sin(\phi ))^{2}]$, \newline
$m_{15}=cos(\theta ))^{2}\left[ \frac{3}{2}(e^{-\beta E_{8}}+e^{-\beta
E_{14}})+(e^{-\beta E_{13}}+e^{-\beta E_{16}})\right] $, \newline
$m_{16}=(e^{-\beta E_{9}}+e^{-\beta E_{15}})[(a_{9}cos(\theta
))^{2}+(c_{9}sin(\theta ))^{2}]$, \newline
\noindent and for the electrons $c$ we have

\begin{equation}
\mathcal{G}_{\alpha\alpha^{\prime}}^{cc,at}(i\omega_{s})=-e^{\beta\Omega}%
\sum_{n,r,r^{\prime}}\frac{\exp(-\beta\varepsilon_{n-1,r})+\exp(-\beta
\varepsilon_{n,r^{\prime}})}{i\omega_{s}+\varepsilon_{n-1,r}-\varepsilon
_{n,r^{\prime}}}\times  \notag
\end{equation}
\begin{equation}
\times \left\langle n-1,r\right\vert\ c_{\vec{k},\alpha}\left\vert
n,r^{\prime}\right\rangle \left\langle n,r^{\prime}\right\vert \ c_{\vec{k}%
,\alpha ^{\prime}}^{\dagger}\ \left\vert n-1,r\right\rangle ,
\label{Eq3.11a}
\end{equation}

\begin{equation}  \label{GAndersona}
G^{cc,at}(\omega) = e^{\beta\Omega} \sum_{i=1}^{16} \frac{n_{i}}{\omega-u_{i}%
},
\end{equation}

\noindent and the residues for the conduction electrons are given by \newline

\noindent $n_{1}=sin(\phi )^{2}[e^{-\beta E_{1}}+e^{-\beta E_{2}}+\frac{3}{2}%
e^{-\beta E_{4}}+\frac{3}{2}e^{-\beta E_{8}}]$, \newline
$n_{2}=cos(\phi )^{2}[e^{-\beta E_{1}}+e^{-\beta E_{5}}+\frac{3}{2}e^{-\beta
E_{3}}+\frac{3}{2}e^{-\beta E_{8}}]$, \newline
$n_{3}=(e^{-\beta E_{2}}+e^{-\beta E_{10}})[(a_{10}cos(\phi
))^{2}+(c_{10}sin(\phi ))^{2}]$, \newline
$n_{4}=(e^{-\beta E_{2}}+e^{-\beta E_{11}})[(a_{11}cos(\phi
))^{2}+(c_{11}sin(\phi ))^{2}]$, \newline
$n_{5}=(e^{-\beta E_{2}}+e^{-\beta E_{9}})[(a_{9}cos(\phi
))^{2}+(c_{9}sin(\phi ))^{2}]$, \newline
$n_{6}=(e^{-\beta E_{4}}+e^{-\beta E_{10}})[(a_{10}sin(\phi
))^{2}+(c_{10}cos(\phi ))^{2}]$, \newline
$n_{7}=cos(\theta )^{2}[e^{-\beta E_{16}}+e^{-\beta E_{14}}+\frac{3}{2}%
e^{-\beta E_{8}}+\frac{3}{2}e^{-\beta E_{13}}]$, \newline
$n_{8}=(e^{-\beta E_{9}}+e^{-\beta E_{12}})[(a_{9}cos(\theta
))^{2}+(b_{9}sin(\theta ))^{2}]$, \newline
$n_{9}=(e^{-\beta E_{10}}+e^{-\beta E_{12}})[(a_{10}cos(\theta
))^{2}+(b_{10}sin(\theta ))^{2}]$, \newline
$n_{10}=(e^{-\beta E_{11}}+e^{-\beta E_{12}})[(a_{11}cos(\theta
))^{2}+(b_{11}sin(\theta ))^{2}]$, \newline
$n_{11}=(e^{-\beta E_{10}}+e^{-\beta E_{14}})[(a_{10}sin(\theta
))^{2}+(b_{10}cos(\theta ))^{2}]$, \newline
$n_{12}=(e^{-\beta E_{11}}+e^{-\beta E_{14}})[(b_{11}cos(\theta
))^{2}+(a_{11}sin(\theta ))^{2}]$. \newline
$n_{13}=(e^{-\beta E_{4}}+e^{-\beta E_{9}})[(c_{9}cos(\phi
))^{2}+(a_{9}sin(\phi ))^{2}]$, \newline
$n_{14}=(e^{-\beta E_{4}}+e^{-\beta E_{11}})[(c_{11}cos(\phi
))^{2}+(a_{11}sin(\phi ))^{2}]$, \newline
$n_{15}=sin(\theta )^{2}[e^{-\beta E_{12}}+e^{-\beta E_{16}}+\frac{3}{2}%
e^{-\beta E_{14}}+\frac{3}{2}e^{-\beta E_{6}}]$, \newline
$n_{16}=(e^{-\beta E_{9}}+e^{-\beta E_{14}})[(b_{9}cos(\theta
))^{2}+(a_{9}sin(\theta ))^{2}]$, \newline

Finally following the same definition for the cumulants in Eq. (\ref{E5.97})
we can write all the atomic Green's functions employed in the calculation of
the atomic approach.

\begin{equation}
\mathbf{G}_{\uparrow }^{ff,at}=%
\begin{pmatrix}
g_{11} & g_{13} \\ 
g_{31} & g_{33}%
\end{pmatrix}%
\hspace{20pt}\mathrm{;\hspace{20pt}}\mathbf{G}_{\downarrow }^{ff,at}=%
\begin{pmatrix}
g_{22} & g_{24} \\ 
g_{42} & g_{44}%
\end{pmatrix}%
,  \label{E55.97}
\end{equation}

\begin{widetext}
\
\begin{eqnarray}
&&g_{11}=e^{-\beta \Omega }\biggl\{\left[ sin^{2}\phi \left( \frac{e^{-\beta
E_{1}}+e^{-\beta E_{4}}}{i\omega +E_{1}-E_{4}}+\frac{3}{2}\frac{e^{-\beta
E_{2}}+e^{-\beta E_{6}}}{i\omega +E_{2}-E_{6}}\right) \right. \newline
\notag \\
&&\left. +\cos ^{2}\phi \left( \frac{e^{-\beta E_{1}}+e^{-\beta E_{2}}}{%
i\omega +E_{1}-E_{2}}+\frac{3}{2}\frac{e^{-\beta E_{4}}+e^{-\beta E_{6}}}{%
i\omega +E_{4}-E_{6}}\right) \right] \newline
\notag \\
&&+\sum_{i=9}^{11}\left[ \frac{e^{-\beta E_{3}}+e^{-\beta E_{i}}}{i\omega
+E_{3}-E_{i}}(a_{i}\sin \phi )^{2}+\frac{e^{-\beta E_{5}}+e^{-\beta E_{i}}}{%
i\omega +E_{5}-E_{i}}(a_{i}\cos \phi )^{2}\right. \newline
\newline
\notag \\
&&\left. +\frac{e^{-\beta E_{i}}+e^{-\beta E_{14}}}{i\omega +E_{i}-E_{14}}%
(c_{i}\sin \theta )^{2}+\frac{e^{-\beta E_{i}}+e^{-\beta E_{12}}}{i\omega
+E_{i}-E_{12}}(c_{i}\cos \theta )^{2}\right] \biggr\},
\end{eqnarray}

\begin{eqnarray}
g_{33} &=&e^{-\beta \Omega }\biggl\{\left[ sin^{2}\theta \left( \frac{%
e^{-\beta E_{15}}+e^{-\beta E_{16}}}{i\omega +E_{15}-E_{16}}+\frac{3}{2}%
\frac{e^{-\beta E_{8}}+e^{-\beta E_{12}}}{i\omega +E_{8}-E_{12}}\right)
\right. \newline
\notag \\
&&\left. +\cos ^{2}\theta \left( \frac{e^{-\beta E_{13}}+e^{-\beta E_{16}}}{%
i\omega +E_{13}-E_{16}}+\frac{3}{2}\frac{e^{-\beta E_{8}}+e^{-\beta E_{14}}}{%
i\omega +E_{8}-E_{14}}\right) \right]  \notag \\
&&+\sum_{i=9}^{11}\left[ \frac{e^{-\beta E_{4}}+e^{-\beta E_{i}}}{i\omega
+E_{4}-E_{i}}(b_{i}\sin \phi )^{2}+\frac{e^{-\beta E_{3}}+e^{-\beta E_{i}}}{%
i\omega +E_{3}-E_{i}}(b_{i}\cos \phi )^{2}\right.  \notag \\
&&\left. +\frac{e^{-\beta E_{i}}+e^{-\beta E_{12}}}{i\omega +E_{i}-E_{12}}%
(a_{i}\sin \theta )^{2}+\frac{e^{-\beta E_{i}}+e^{-\beta E_{14}}}{i\omega
+E_{i}-E_{14}}(a_{i}\cos \theta )^{2}\right] \biggr\},
\end{eqnarray}

\begin{eqnarray}
g_{13} &=&e^{-\beta \Omega }\sum_{i=9}^{11}\biggl\{\left[ \frac{e^{-\beta
E_{5}}+e^{-\beta E_{i}}}{i\omega +E_{5}-E_{i}}-\frac{e^{-\beta
E_{3}}+e^{-\beta E_{i}}}{i\omega +E_{3}-E_{i}}\right] (a_{i}b_{i}\sin \phi
\cos \phi )  \notag \\
&&+\left[ \frac{e^{-\beta E_{i}}+e^{-\beta E_{14}}}{i\omega +E_{i}-E_{14}}-%
\frac{e^{-\beta E_{i}}+e^{-\beta E_{12}}}{i\omega +E_{i}-E_{12}}\right]
(a_{i}c_{i}\sin \theta \cos \theta )\biggr\}
\end{eqnarray}

\begin{equation}
g_{31}= g_{13} ,
\end{equation}

\begin{eqnarray}
g_{22} &=&e^{-\beta \Omega }\biggl\{\left[ sin^{2}\phi \left( \frac{%
e^{-\beta E_{1}}+e^{-\beta E_{5}}}{i\omega +E_{1}-E_{5}}+\frac{3}{2}\frac{%
e^{-\beta E_{3}}+e^{-\beta E_{7}}}{i\omega +E_{3}-E_{7}}\right) \right.
\newline
\notag \\
&&\left. +\cos ^{2}\phi \left( \frac{e^{-\beta E_{1}}+e^{-\beta E_{3}}}{%
i\omega +E_{1}-E_{3}}+\frac{3}{2}\frac{e^{-\beta E_{5}}+e^{-\beta E_{7}}}{%
i\omega +E_{5}-E_{7}}\right) \right]   \notag \\
&&+\sum_{i=9}^{11}\left[ \frac{e^{-\beta E_{2}}+e^{-\beta E_{i}}}{i\omega
+E_{2}-E_{i}}(a_{i}\sin \phi )^{2}+\frac{e^{-\beta E_{4}}+e^{-\beta E_{i}}}{%
i\omega +E_{4}-E_{i}}(a_{i}\cos \phi )^{2}\right.   \notag \\
&&\left. +\frac{e^{-\beta E_{i}}+e^{-\beta E_{15}}}{i\omega +E_{i}-E_{15}}%
(c_{i}\sin \theta )^{2}+\frac{e^{-\beta E_{i}}+e^{-\beta E_{13}}}{i\omega
+E_{i}-E_{13}}(c_{i}\cos \theta )^{2}\right] \biggr\},
\end{eqnarray}

\begin{eqnarray}
g_{44} &=&e^{-\beta \Omega }\biggl\{\left[ sin^{2}\theta \left( \frac{%
e^{-\beta E_{14}}+e^{-\beta E_{16}}}{i\omega +E_{14}-E_{16}}+\frac{3}{2}%
\frac{e^{-\beta E_{6}}+e^{-\beta E_{12}}}{i\omega +E_{6}-E_{12}}\right)
\right. +  \notag \\
&&\left. +\cos ^{2}\theta \left( \frac{e^{-\beta E_{12}}+e^{-\beta E_{16}}}{%
i\omega +E_{12}-E_{16}}+\frac{3}{2}\frac{e^{-\beta E_{6}}+e^{-\beta E_{14}}}{%
i\omega +E_{6}-E_{14}}\right) \right]   \notag \\
&&\sum_{i=9}^{11}\left[ \frac{e^{-\beta E_{4}}+e^{-\beta E_{i}}}{i\omega
+E_{4}-E_{i}}(b_{i}\sin \phi )^{2}+\frac{e^{-\beta E_{2}}+e^{-\beta E_{i}}}{%
i\omega +E_{2}-E_{i}}(b_{i}\cos \phi )^{2}\right.   \notag \\
&&\left. +\frac{e^{-\beta E_{i}}+e^{-\beta E_{13}}}{i\omega +E_{i}-E_{13}}%
(a_{i}\sin \theta )^{2}+\frac{e^{-\beta E_{i}}+e^{-\beta E_{15}}}{i\omega
+E_{i}-E_{15}}(a_{i}\cos \theta )^{2}\right] \biggr\},
\end{eqnarray}

\begin{eqnarray}
g_{24} &=&e^{-\beta \Omega }\sum_{i=9}^{11}\biggl\{\left[ \frac{e^{-\beta
E_{2}}+e^{-\beta E_{i}}}{i\omega +E_{2}-E_{i}}-\frac{e^{-\beta
E_{4}}+e^{-\beta E_{i}}}{i\omega +E_{4}-E_{i}}\right] (a_{i}b_{i}\sin \phi
\cos \phi )  \notag \\
&&+\left[ \frac{e^{-\beta E_{i}}+e^{-\beta E_{13}}}{i\omega +E_{i}-E_{13}}-%
\frac{e^{-\beta E_{i}}+e^{-\beta E_{15}}}{i\omega +E_{i}-E_{15}}\right]
(a_{i}c_{i}\sin \theta \cos \theta )\biggr\},
\end{eqnarray}

\begin{equation}
g_{42}= g_{24} .  \label{E55.99}
\end{equation}

\end{widetext}

\end{document}